\documentclass[twocolumn]{aastex631}

\usepackage{times,amsmath}
\usepackage[T1]{fontenc}

\hypersetup{pdfauthor={},
            pdftitle={},
            pdfkeywords={},
            bookmarksnumbered=true}
\pdfoutput=1

\usepackage[utf8]{inputenc}
\usepackage{graphicx}
\usepackage{amssymb}
\usepackage{amsmath}
\usepackage{float}
\usepackage{multirow}
\usepackage{textcomp}
\usepackage{gensymb}
\usepackage{enumitem}  
\usepackage{hyperref}{
\usepackage{natbib}
\usepackage{comment}
\usepackage{systeme}
\usepackage[Symbol]{upgreek}
\usepackage{xcolor}
\usepackage{xspace}
\definecolor{xlinkcolor}{cmyk}{1,1,0,0}

\newcommand{\s}{\ensuremath{\mbox{~s}}}

\newcommand{\gtsim}{\,\rlap{\raise 0.5ex\hbox{$>$}}{\lower 1.0ex\hbox{$\sim$}}\,}

\shorttitle{Radio Halos Over 2~Mpc in Massive Galaxy Clusters }
\shortauthors{ Rajpurohit et al.}

\begin{document}

\title{Radial Profiles of Radio Halos in Massive Galaxy Clusters: Diffuse Giants Over 2 Mpc}

\correspondingauthor{Kamlesh Laxmi Rajpurohit}
\email{kamlesh.rajpurohit@cfa.harvard.edu}

\author[0000-0001-7509-2972]{K. Rajpurohit}
\affiliation{Center for Astrophysics $|$ Harvard \& Smithsonian, 60 Garden
  Street, Cambridge, MA 02138, USA}
\affil{INAF-IRA, via Gobetti 101, 40129 Bologna, Italy} 

\author[0000-0002-9325-1567]{A. Botteon}
\affil{INAF-IRA, via Gobetti 101, 40129 Bologna, Italy} 

\author[0000-0002-5671-6900]{E. O'Sullivan}
\affil{Center for Astrophysics $|$ Harvard \& Smithsonian, 60 Garden Street, Cambridge, MA 02138, USA}

\author[0000-0002-9478-1682]{W. Forman}
\affil{Center for Astrophysics $|$ Harvard \& Smithsonian, 60 Garden Street, Cambridge, MA 02138, USA}

\author{M. Balboni}
\affil{INAF-IASF Milano, via A. Corti 12, 20133 Milano, Italy}
\affil{DIFA - Universit\`a di Bologna, via Gobetti 93/2, I-40129 Bologna, Italy}

\author{L. Bruno}
\affil{INAF-IRA, via Gobetti 101, 40129 Bologna, Italy} 

\author[0000-0002-0587-1660]{R. J. van Weeren}
\affiliation{Leiden Observatory, Leiden University, PO Box 9513, 2300 RA Leiden, The Netherlands}

\author{M. Hoeft}
\affiliation{Th\"{u}ringer Landessternwarte, Sternwarte 5, 07778 Tautenburg, Germany}

\author{G. Brunetti}
\affil{INAF-IRA, via Gobetti 101, 40129 Bologna, Italy} 

\author[0000-0003-2206-4243]{C. Jones}
\affil{Center for Astrophysics $|$ Harvard \& Smithsonian, 60 Garden Street, Cambridge, MA 02138, USA}

\author{A. S. Rajpurohit}
\affil{Astronomy \& Astrophysics Division, Physical Research Laboratory, Ahmedabad 380009, India}

\author{S. P. Sikhosana}
\affiliation{Astrophysics Research Centre, University of KwaZulu-Natal, Durban, 3696, South Africa}

\begin{abstract}
We present new, high frequency radio observations of the merging galaxy clusters PLCK\,G287.0+32.9, Abell 2744, and Bullet. These clusters are known to host $\sim$Mpc scale sources, known as radio halos, which are formed by the acceleration of cosmic rays by turbulence injected into the intracluster medium during cluster mergers. Our new images reveal previously undetected faint outermost regions of halos, extending to over 2~Mpc. This discovery highlights the presence of radio halos with large extents at high frequencies and suggests that their observable size depends on a combination of the observation sensitivity and uv-coverage, and their radio power. We additionally compare the properties of these three clusters with MACS\,J0717+3745 and Abell 2142, both of which are known to host prominent large radio halos. Remarkably, all five halos, despite their exceptionally large extents, exhibit properties similar to other classical halos: their radial profiles are described by a single-component exponential fit, they show radial spectral index steepening, and have an average radio emissivity of about $10^{-42}\, \mathrm{erg\,s^{-1}\,cm^{-3}\,Hz^{-1}}$. Our results demonstrate that radio halos can extend to the cluster periphery, without the transition to an observationally distinguishable different halo component in the outermost regions. Our findings also highlight that careful subtraction of unrelated sources embedded in the halo is necessary to measure the radio surface brightness accurately, as incomplete subtraction can introduce an apparent secondary component in the peripheral regions.

\end{abstract}


\section{Introduction}
Radio halos are among the largest class of extended diffuse radio sources in the intracluster medium (ICM). They are found at the center of merging clusters and typically correlate with the X-ray emission \citep[e.g.,][]{Govoni2001a,Rajpurohit2018, Botteon2020, Bonafede2022}. They are believed to be formed by turbulent re-acceleration of cosmic ray electrons (CRe) which become radio-emitting by turbulence, injected into the ICM by cluster mergers \citep{Brunetti2001, Petrosian2001, Brunetti2007, Brunetti2014,Miniati2015, vanWeeren2019}. Secondary particles, produced via hadronic collisions in the ICM, may provide an additional mechanism to generate CRe in clusters \citep[e.g.,][]{Blasi1999, Dolag2000, Pfrommer2008}. However, the current gamma-ray limits disfavor this scenario \citep{Pinzke2017, Brunetti2017}. 

\begin{figure}[!thbp]
    \centering
    \includegraphics[width=0.46\textwidth]{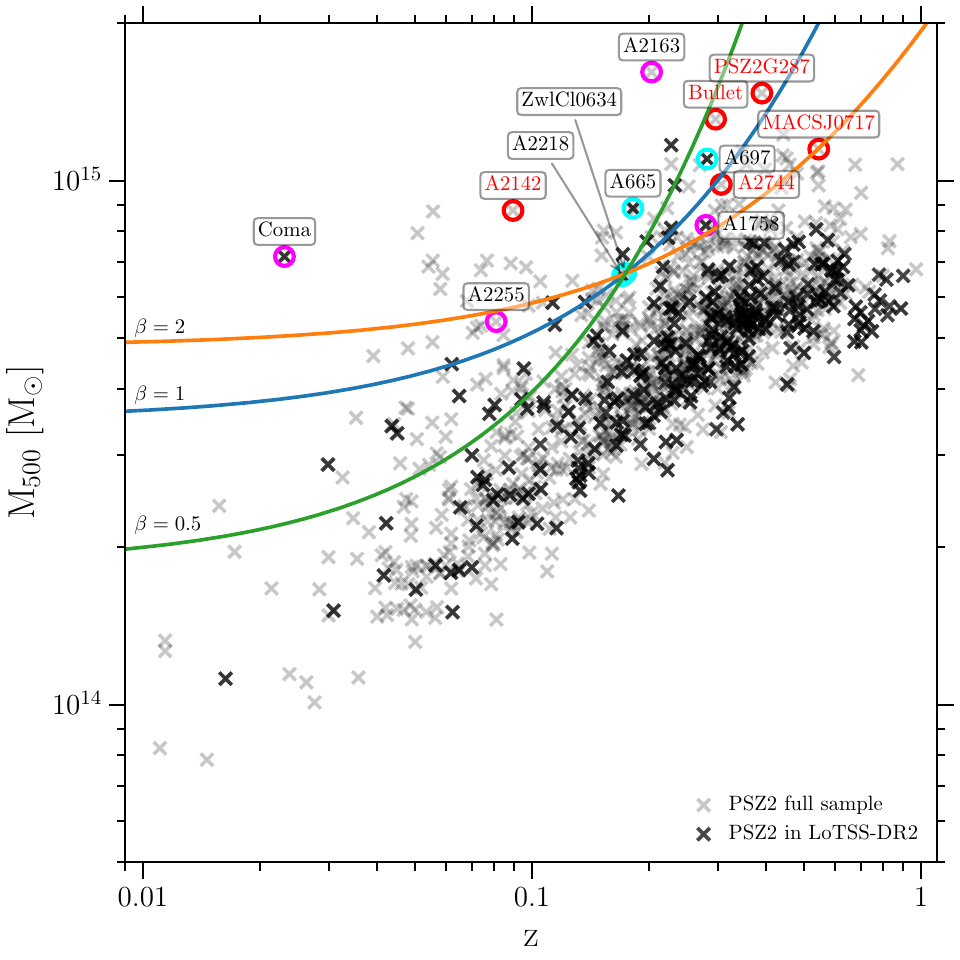}
 \caption{Cluster mass versus redshift distribution of the PSZ2 clusters. The subsample of PSZ2 in LoTSS-DR2 \citep{Botteon2022a} is reported in black. Circles denote clusters with halos having LLS $>$ 2\,Mpc and are color-coded as follows: in cyan are the mega-halos  \citep{Cuciti2022}, in red are the clusters analyzed in this work, and in magenta are other clusters reported in the literature. The solid lines represent the boundaries above which mega-halos are assumed to be detected with LOFAR observations assuming that their surface brightness is $\propto M_{500}^\beta (1+z)^{-4}$, for three possibilities for $\beta$.}
      \label{M_z_plot}
\end{figure}

The remarkable developments in radio sensitivity and resolution of modern instruments are uncovering larger extents of these diffuse radio halos along with intricate embedded structures \citep[e.g.,][]{Knowles2022, Botteon2022a, Botteon2022, Duchesne2024, Botteon2024, Rajpurohit2023, Rajpurohit2021c, vanWeeren2017b, vanWeeren2016a, Osinga2024, Sikhosana2023}. In particular, while a few radio halos with very large linear size (LLS), exceeding 2 Mpc, were already reported with ``old generation'' interferometers \cite[e.g. Bullet and Abell 2163][]{Feretti2001, Liang2000}. Radio halos with such extents are being detected more frequently in a growing number of systems, see Figure\,\ref{M_z_plot} \citep{Sikhosana2023,Bruno2023,Rajpurohit2021b, Rajpurohit2021c, Bonafede2022, Shweta2020}. With the deepest low-frequency observations carried out on a cluster with LOFAR, for example, a total extent of approximately 5~Mpc for the diffuse emission in Abell 2255 has been found \citep{Botteon2022}. Additionally, even at high frequencies, diffuse emission, extending approximately 6~Mpc, is observed in PLCK\, G287.0+32.9 (Rajpurohit et al. in prep.).

Recently, by analyzing a sample of $\sim310$ Planck clusters, \citet{Cuciti2022} suggested the existence of a new class of radio sources, called mega-halos, based on the detection of diffuse emission in four clusters that exhibit properties distinct from those typically observed in classical radio halos.  These sources were detected with LOFAR (at low resolution) and were found to be hosted in massive galaxy clusters (see Figure\,\ref{M_z_plot}). Their defining characteristics include an LLS exceeding 2\,Mpc, extending at least up to $R_{500}$ (the radius within which the average mass density of a cluster is 500 times the critical density of the Universe), and a two-component radial radio surface brightness profile. The first component appears as a brighter, canonical halo-like structure following an exponential profile, while the second component is reported to be shallower. Additionally, based on the spectral analysis performed at 50-150~MHz for two clusters, the outer component of the mega-halo is reported to exhibit an ultra-steep spectrum ($\alpha\geq -1.5$) 

The presence of two components in the radial surface brightness profiles has also been observed in relaxed clusters hosting mini-halo (related to sloshing motions) and halo like emission  \citep[e.g.,][]{Biava2024, vanWeeren2024}, dubbed hybrid halos, which typically show LLS below 1~Mpc. In these systems, the outer, fainter component exhibits characteristics typical of classical halos. Moreover, some classical radio halos also exhibit multiple components within the halo emission \citep{Rajpurohit2021c, Bruno2023, Rajpurohit2023}. Whether these two-component systems (halos and mega-halos) constitute a new class of sources that lie below or encompass classical radio halos remains an open question, requiring deeper multi-frequency observations and larger sample studies.

The main goal of this paper is to investigate the observational properties of radio halos, where the new observations reveal additional extents of their diffuse emission. Using high-quality observations, we analyzed radial profiles of the following five clusters: PLCK\,G287.0+32.9, Abell 2744, the Bullet Cluster, MACS\,J0717+3745, and Abell 2142. These targets were selected due to their high mass and availability of deep, multifrequency radio observations. We note that to date mega-halos are detected in the redshift range of $z\sim0.17-0.28$ and mass $M_{500} \sim 6-11\times 10^{14}M_{\odot}$.  Their surface brightness is assumed to scale with $\propto M_{500}^\beta (1+z)^{-4}$. In Figure\,\ref{M_z_plot}, we show the three possibilities for $\beta$. PLCK\,G287+32.9, Abell 2744, the Bullet Cluster, MACS\,J0717+3745, and Abell 2142 are ideal clusters to host mega-halos as supported by their location in the $M-z$ distribution. We investigate whether or not these halos with large extents exhibit characteristics typical of classical halos.   

\setlength{\tabcolsep}{15pt}
\begin{table*}
\caption{Observational overview of PLCK\,G287.0+32.9.}
\begin{center}
\begin{tabular}{ l  c c c c}
  \hline  \hline  
& \multicolumn{3}{c}{ MeerKAT} & \multirow{1}{*}{uGMRT} \\  
 \cline{2-4}
 & \multirow{1}{*}{UHF} &\multirow{1}{*}{L-band}& \multirow{1}{*}{S-band} &\multirow{1}{*}{Band3}\\
\hline
Observing date & April 1, 2023 & February 20, 2023 &August 24, 2024 & December 31, 2023 \\
Frequency coverage &0.5-1.0\,GHz& 0.9-1.7\,GHz&1.9-2.8\,GHz&300-500\,MHz \\
Channel width  &132.81\,kHz & 208\,kHz&213\,kHz&42\,kHz\\ 
Number of channels  &4096 & 4096&4096&4096\\ 
On source time &5\,hrs &6\,hrs&3\,hr&8\,hrs\\
\hline 
\end{tabular}
\end{center}
\label{Tabel:obs}
\end{table*}

\setlength{\tabcolsep}{3pt}
\begin{table}
\caption{Observational overview of Abell 2744.}
\begin{center}
\begin{tabular}{ l  l l c c}
  \hline  \hline  
&  \multirow{1}{*}{MeerKAT} & \multirow{1}{*}{uGMRT} \\  
 & \multirow{1}{*}{UHF} &\multirow{1}{*}{Band4}\\
\hline
Observing date & October 29, 2023 & September 5-6, 2019 \\
 & October 31, 2023 & September 11-14, 2019 \\
 & November 1, 2023 & September 21, 2019 \\
Frequency coverage &0.5-1.0\,GHz&550-850\,MHz \\
Channel width  &132.81\,kHz & 97.7\,kHz\\ 
Number of channels  &4096 &4096\\ 
On source time &12\,hrs &36\,hrs\\
\hline 
\end{tabular}
\end{center}
\label{Tabel:obs1}
\end{table}

\setlength{\tabcolsep}{6pt}
\begin{table}
\caption{Observational overview of Bullet cluster.}
\begin{center}
\begin{tabular}{ l  l l c c}
  \hline  \hline  
&  \multicolumn{2}{c}{MeerKAT}  \\  
 \cline{2-3}
 & \multirow{1}{*}{UHF} &\multirow{1}{*}{L-band}\\
\hline
Observing date & January 29-30, 2025 & June 24, 2018 \\
& December 13-14, 2024 &\\
Frequency coverage &0.5-1.0\,GHz&0.9-1.7\,GHz \\
Channel width  &132.81\,kHz &23.3\,kHz \\ 
Number of channels  &4096 &4096\\ 
On source time &8\,hrs &8\,hrs\\
\hline 
\end{tabular}
\end{center}
\label{Tabel:obs3}
\end{table}

Throughout this paper, we adopt a flat $\Lambda$CDM cosmology with $H_{\rm{ 0}}=70$ km s$^{-1}$\,Mpc$^{-1}$, $\Omega_{\rm{ m}}=0.30$, and $\Omega_{\Lambda}=0.71$. We define the spectral index, $\alpha$, so that $S_{\nu}\propto\nu^{\alpha}$, where $S$ is the flux density at frequency $\nu$ ($\alpha\leq-1$). All output radio images are in the J2000 coordinate system. The LLS and radial profiles of all radio halos presented in this study are measured at $\geq 3\sigma_{\rm rms}$ level, where $\sigma_{\rm rms}$ is the noise level.

\vspace{-1.2em}
\begin{deluxetable*}{c c c c r c c c r}[!thbp]
\tablecaption{Imaging properties of radio maps used in the analysis}
\tablehead{Cluster& Central Frequency  &Image Name & Restoring Beam & Robust  & \textit{uv}-cut & \textit{uv}-taper & RMS noise\\ 
&GHz &&& parameter &&&$\upmu\rm Jy\,beam^{-1}$}
\startdata
\multirow{2}{3cm}{PLCK\,G287.0+32.9} &2.40& IM1&$20\arcsec \times 20\arcsec$&$0.0$&$-$&$15\arcsec$&8\\
 & 1.28&IM2&$20\arcsec \times 20\arcsec$&$0.0$&$-$&15\arcsec&15\\
 & 0.815 &IM3&$20\arcsec \times 20\arcsec$&$0.0$&$-$&15\arcsec& 25\\
 & 0.400&IM4&$20\arcsec \times 20\arcsec$&$0.0$&$-$&15\arcsec&  32\\
 &2.40&IM5&$50\arcsec \times 50\arcsec$&$-0.5$&$\geq\rm0.2\,k\uplambda$&$40\arcsec$&13\\
 & 1.28&IM6&$50\arcsec \times 50\arcsec$&$-0.5$&$\geq\rm0.2\,k\uplambda$&40\arcsec&25\\
 &  0.815&IM7&$50\arcsec \times 50\arcsec$&$-0.5$&$\geq\rm0.2\,k\uplambda$&40\arcsec& 40 \\
 & 0.400&IM8&$50\arcsec \times 50\arcsec$&$-0.5$&$\geq\rm0.2\,k\uplambda$&40\arcsec&  100\\
\hline
 \multirow{3}{3cm}{Abell 2744}&0.815 &IM9&$25\arcsec \times 20\arcsec$& $0.0$&$-$&$20\arcsec$&24\\
  & 0.675&IM10&$25\arcsec \times 25\arcsec$&$0.0$&$ -$&20\arcsec&20\\
   &1.50&IM11&$25\arcsec \times 25\arcsec$& $-0.5$&$\geq\rm0.2\,k\uplambda$&$20\arcsec$&23\\
  &0.815&IM12&$25\arcsec \times 25\arcsec$& $-0.5$&$\geq\rm0.2\,k\uplambda$&$20\arcsec$&28\\
 &0.675&IM13&$25\arcsec \times 25\arcsec$&$-0.5$&$ \geq\rm0.2\,k\uplambda$&20\arcsec&25\\ 
 \hline   
\multirow{3}{3cm}{Bullet} &1.28&IM14&$20\arcsec \times 20\arcsec$&$0.0$&$-$&$15\arcsec$&18 \\
&0.815&IM15&$20\arcsec \times 20\arcsec$&$0.0$&$ -$&15\arcsec&20 \\
& 1.28&IM16&$20\arcsec \times 20\arcsec$&$-0.5$&$\geq\rm0.2\,k\uplambda$&$15\arcsec$&20 \\
&0.815 &IM17&$20\arcsec \times 20\arcsec$&$-0.5$&$\geq\rm0.2\,k\uplambda$&$15\arcsec$& 22\\
 \hline     
\multirow{2}{3cm}{MACSJ0717}& 0.400&IM18&$20\arcsec \times 20\arcsec$&$-0.5$ &$\geq\rm0.2\,k\uplambda$&$15\arcsec$&180\\
 &0.144&IM19&$20\arcsec \times 20\arcsec$&$-0.5$&$\geq\rm0.2\,k\uplambda$&$15\arcsec$& 40 \\
\enddata
\tablecomments{Final imaging was performed in \texttt{WSCLEAN} using {\tt multiscale} and with the {\tt Briggs} weighting scheme. For the Abell 2142 image properties, we refer to \cite{Bruno2023} }
\label{tab:imaging}
\end{deluxetable*}

\section{Observations and data reduction}
We analyzed new observations of PLCK\,G287.0+32.9, Abell\,2744, and the Bullet Cluster, along with published data for MACS\,J0717.5+3745 and Abell\,2142. For details on the observations and data reduction of MACS\,J0717.5+3745 and Abell 2142, we refer to \citet{Rajpurohit2021a, Rajpurohit2021b, Bruno2023}. 

\subsection{PLCK\,G287.0+32.9 }
\label{PSZ_datareduction}
The cluster was observed with MeerKAT in the UHF 0.55-1.0 GHz (Project code: SCI-20220822), L-band 1-1.7~GHz (Project code: SCI-20220822) and S-band 1.9-2.8 GHz (Project code: SCI-20230907),  see Table\,\ref{Tabel:obs} for details.  All four polarization products were recorded using the 4K correlator mode, covering a total bandwidth 544\,MHz and 875\,MHz at UHF and S-band, respectively.  For the UHF and S-band observations, J0408-6545 was the primary calibrator used for flux and bandpass calibration, observed at the beginning and end of the observing run. J1154-3505 was observed as a gain calibrator and J0521+1638 as a polarization calibrator.

The MeerKAT data were calibrated using the Containerized Automated Radio Astronomy Calibration \citep[$\tt{CARACal}$;][]{caracal2020} pipeline {\footnote{\url{https://ascl.net/2006.014}}}. The first step consists of flagging in {\tt CARACAL}, including shadowed antennas, autocorrelations, known RFI channels, and the tfcrop algorithm. Thereafter, {\tt AOflagger} \citep{Offringa2010} was used to flag bad data using the CARACal {\tt firstpass\_QUV.rfis} strategy.  {\tt CARACal} modeled the primary calibrator J0408-6545 using the MeerKAT local sky models. Following this, cross-calibration was performed to solve for the time-dependent delays and complex gains of each antenna and the bandpass corrections.

After initial calibration, we flagged the remaining low surface brightness RFI using {\tt AOflagger}. We then created an initial image of the target field using {\tt WSClean} \citep{Offringa2014} within {\tt CARACAL}. Three rounds of phase-only self-calibration were performed using {\tt CubiCal}  \citep{Kenyon2018}, followed by a final round of amplitude-phase calibration. The calibrated data were imaged in  {\tt WSClean} using the Briggs weighting scheme with a robust parameter of $0$, and multiscale cleaning. For the L-band data reduction, we refer to Balboni et al. in prep.  

We also observed the cluster with the upgraded Giant Metrewave Radio Telescope (uGMRT) at Band3 (Project code: 45\_020 ), covering the frequency range of 300-500\,MHz. The uGMRT data calibration was performed using the Source Peeling and Atmospheric Modeling pipeline \citep[\texttt{SPAM};][]{Intema2009}. We first split the wideband dataset into six sub-bands. The flux density of the primary calibrator 3C\,286 was set according to \cite{Scaife2012}. Following this, the data were averaged, flagged, and corrected for the bandpass. We used a global sky model obtained from the GMRT GSB data to correct the phase gains of the target. Finally, the {\tt SPAM} calibrated sub-bands were imaged in {\tt WSClean} to produce deep, full continuum images using Briggs weighting scheme with a robust parameter of $0$, and multiscale cleaning.

\begin{figure*}[!thbp]
    \centering
    \includegraphics[width=0.86\textwidth]{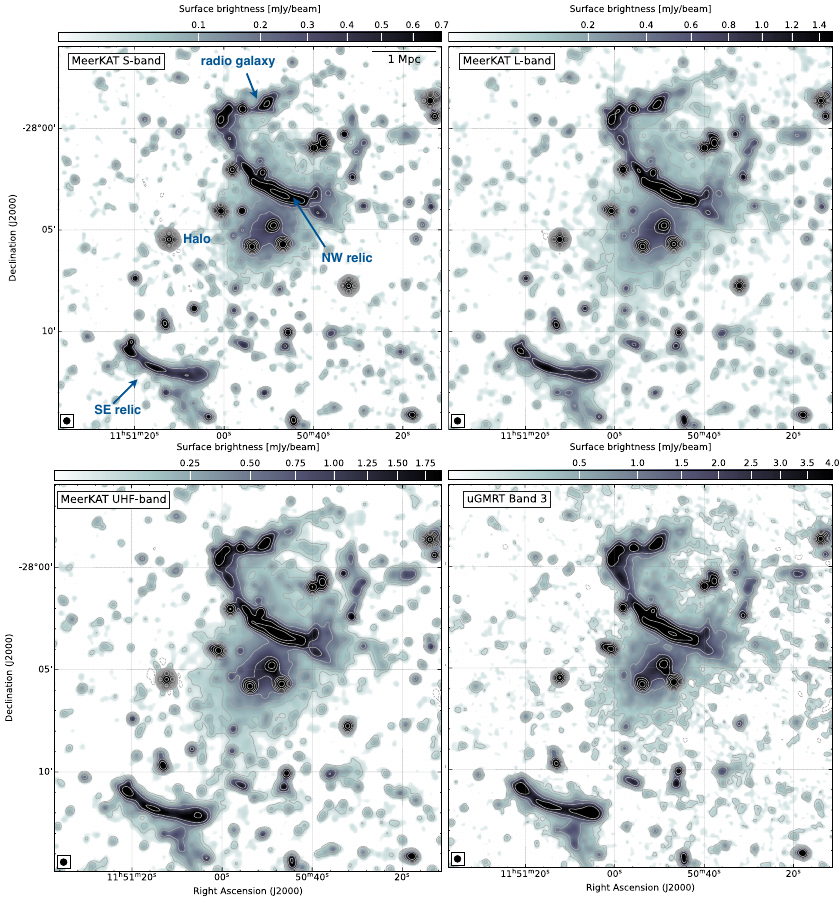}
        \vspace{-0.3cm}
 \caption{PLCK~G287.0+32.9 field (square-root scale) full band continuum images from MeerKAT observations using S-band  (1.9-2.8\,GHz), L-band (0.9-1.7\,GHz) and UHF bands (0.5-1.0\,GHz) and uGMRT Band3 (300-500\,MHz). All images have a common resolution of 20\arcsec. The radio beam size is indicated in the bottom left corner of each image. The images show large scale diffuse radio emission including two symmetrically located relics and a 3.5~Mpc radio halo. The contour levels are drawn at  $[1, 2, 4, 8 ...]\times 3\sigma_{\rm rms}$. Dashed contours depict the $-3\sigma_{\rm rms}$ contours. For image properties, see Table\,\ref{tab:imaging} (IM1, IM2, IM3 and IM4).}
      \label{fig::PSZ_halo}
\end{figure*}

\begin{figure*}[!thbp]
    \centering
    \includegraphics[width=0.86\textwidth]{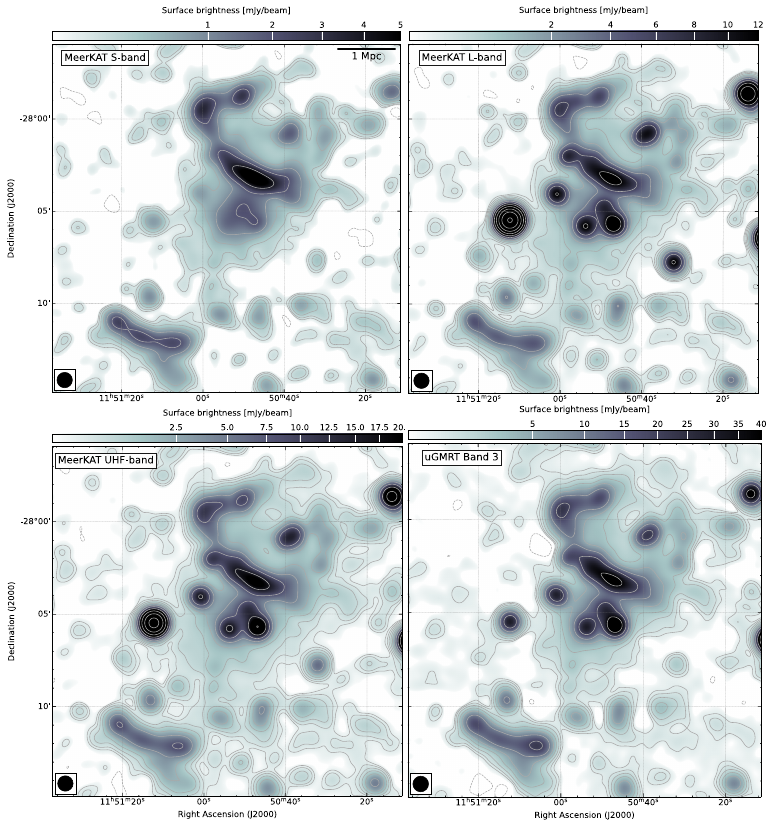}
        \vspace{-0.3cm}
 \caption{PLCK~G287.0+32.9 field (square-root scale) full band continuum images from MeerKAT observations using S-band  (1.9-2.8\,GHz), L-band (0.9-1.7\,GHz) and UHF bands (0.5-1.0\,GHz) and uGMRT Band3 (300-500\,MHz). All images have a common resolution of 50\arcsec. The radio beam size is indicated in the bottom left corner of each image. The images are used to extract radial surface brightness profiles of the halo as a function of frequency. The contour levels are drawn at  $[1, 2, 4, 8 ...]\times 3\sigma_{\rm rms}$. Dashed contours depict the $-3\sigma_{\rm rms}$ contours. For image properties, see Table\,\ref{tab:imaging} (IM5, IM6, IM7 and IM8).}
      \label{fig::PSZ_halo_50arcsec}
\end{figure*}

\begin{figure*}[!thbp]
    \centering
    \includegraphics[width=0.96\textwidth]{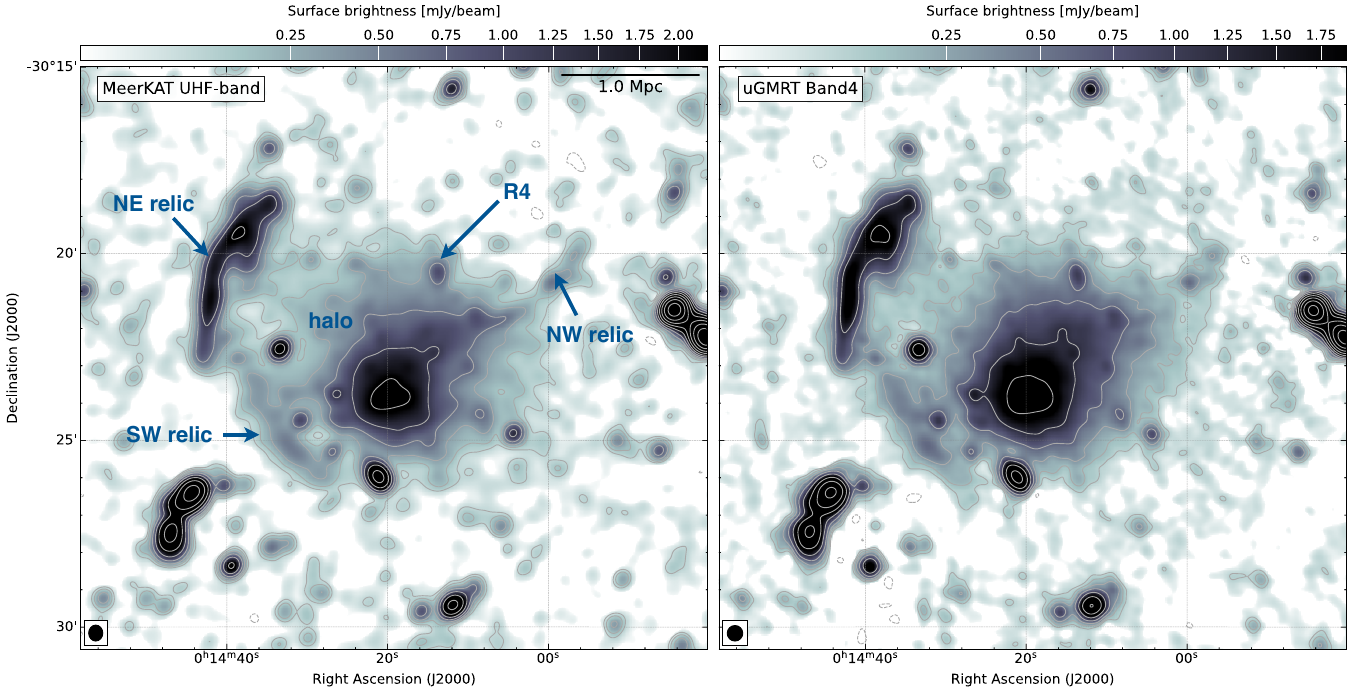}
 \caption{A2744 full band continuum images of the central field from MeerKAT UHF (0.55-1~GHz) and uGMRT Band4 (550-850~MHz) observations (in square-root scale).  The figures show central halo emission and four relics in the outskirts \citep{Rajpurohit2021c}. Both images have a common resolution of 25\arcsec. The radio beam size is indicated in the bottom left corner of each image. The contour levels are drawn at  $[1, 2, 4, 8 ...]\times 3\sigma_{\rm rms}$. Dashed contours depict the $-3\sigma_{\rm rms}$ contours. For image properties, see Table\,\ref{tab:imaging} (IM9 and IM10).}
      \label{fig::A2744_halo}

\end{figure*}

\subsection{Abell 2744}
A2744 was observed with MeerKAT in the UHF band (Project code: SCI-20230907), covering a frequency range of 0.55-1.0~GHz. The observations were conducted over three separate observing runs. J0408-6545 was used as the flux calibrator, while J0025-2602 served as the gain calibrator. For observational details, we refer to Table\,\ref{Tabel:obs1}. We followed the same procedure described in Section~\ref{PSZ_datareduction} for data reduction using {\tt CARACAL}. 

For the uGMRT Band4 data reduction, we refer to \cite{Rajpurohit2021c}, as we used the same calibrated data presented in that work. Both MeerKAT and uGMRT calibrated data were imaged in {\tt WSClean} using Briggs weighting scheme with a robust parameter of $0$, and multiscale cleaning.

\subsection{Bullet cluster}

We used calibrated MeerKAT L-band data presented in \cite{Sikhosana2023}. For the UHF-band, we used new MeerKAT observations (Project code: SCI-20241101). The observations were performed over two observing runs. J0408-6545 and J1939-6342 were used as flux calibrators, while J0825-5010 as a gain calibrator. The observational details are summarized in Table~\ref{Tabel:obs3}. The data were processed by the SARAO Science Data Processor pipeline\footnote{\url{https://skaafrica.atlassian.net/wiki/spaces/ESDKB/pages/338723406/SDP+pipelines+overview}}. The calibrated data were imaged in {\tt WSClean} using Briggs weighting scheme with a robust parameter of $0$, and multiscale cleaning.

All images have been corrected for primary beam attenuation using {\tt EveryBeam} within {\tt WSClean}. For MeerKAT data, the primary beam corrected flux density measurements obtained from both {\tt katbeam}\footnote{\url{https://github.com/ska-sa/katbeam}} and {\tt EveryBeam}\footnote{\url{https://everybeam.readthedocs.io/en/latest/}} were found to be consistent.

\begin{figure*}[!thbp]

    \centering
    \includegraphics[width=0.92\textwidth]{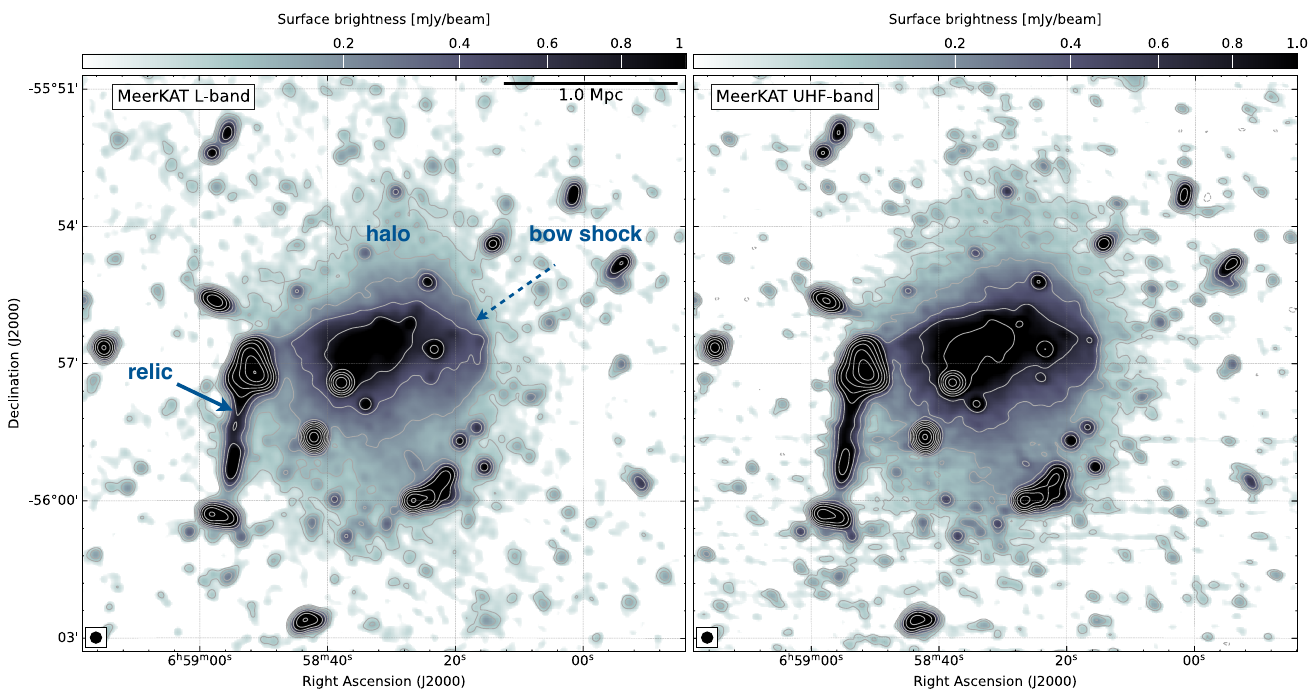}
     \vspace{-0.1cm}  
 \caption{Bullet cluster MeerKAT L-band (0.9-1.7~GHz) and UHF (0.55-1.0~GHz) full band continuum radio images in square-root scale, showing a large scale central halo emission and a toothbrush shaped relic to the east \citep{Sikhosana2023}. Both images have a common resolution of 15\arcsec. The radio beam size is indicated in the bottom left corner of each image. The images show large scale asymmetrical halo emission. The contour levels are drawn at  $[1, 2, 4, 8 ...]\times 3\sigma_{\rm rms}$. Dashed contours depict the $-3\sigma_{\rm rms}$ contours. For image properties, see Table\,\ref{tab:imaging} (IM14 and IM15).}
      \label{fig:Bullet}
\end{figure*}

\section{Radio continuum maps of individual clusters}

Figures\,\ref{fig::PSZ_halo}  and \ref{fig::PSZ_halo_50arcsec} show our new MeerKAT (UHF, L and S bands) and uGMRT Band3 images of PLCK\,G287+32.9. At a redshift of $z = 0.39$, this cluster stands out as an exceptionally luminous and massive (the second most massive)  in the Planck sample  \citep{Planck2016}. A pair of radio relics, a central radio halo, and three filamentary features were previously reported in the GMRT, Very Large Array, and Murchison Widefield Array observations \citep{Bagchi2011, Bonafede2014, George2017}. Our new radio observations revealed 6~Mpc diffuse radio emission filling the entire cluster volume and a multitude of new structures. In this paper, we only focus on the radio halo emission. 

We find that the halo in PLCK\,G287+32.9 is about 3, 3.5, 3.5, and 2.5 Mpc in diameter at 350~MHz, 815~MHz, 1.28 GHz, and 2.4\,GHz, respectively (Figure\,\ref{fig::PSZ_halo_50arcsec}). To the best of our knowledge, this is the first halo with LLS of about 2.5 Mpc at 2.4~GHz. The total extent of the PLCK\,G287+32.9 halo is at least 2.5 times larger than reported earlier \citep{Bonafede2014}. The overall morphology of the halo emission is similar from 350~MHz to 2.4~GHz. However, the halo is more extended toward low frequencies as found in other well-studied halos, for example, 1RXS J0603.3+4214 \citep{vanWeeren2016a, Rajpurohit2020a}, Abell 2744 \citep{Pearce2017, Rajpurohit2021c}, MACSJ0717+3745 \citep{vanWeeren2017b, Bonafede2018, Rajpurohit2021b}, CIZA J2242.8+5301 \citep{Gennaro2018, Hoang2017}. At 350~MHz, we do not recover the faint outermost regions of the PLCK\,G287+32.9 halo observed in the MeerKAT UHF/L-band images. This is due to the superior sensitivity and dense inner uv-coverage of the MeerKAT compared to the uGMRT.

\begin{figure*}[!thbp]
 \vspace{-0.23cm}
    \centering
    \includegraphics[width=0.90\textwidth]{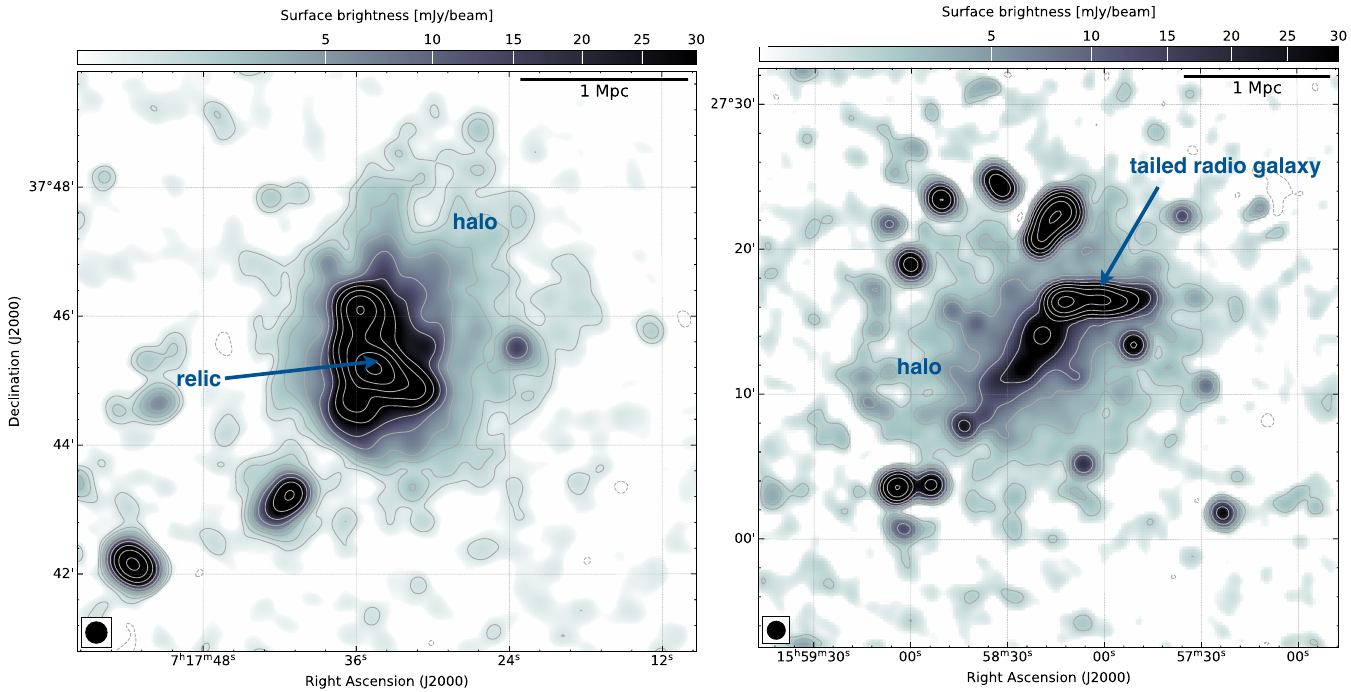}
     \vspace{-0.2cm}
 \caption{ MAC\,J0717 LOFAR HBA (120-167~MHz) and Abell 2142 (120-167~MHz) full band continuum radio images in square-root scale. The central region of MAC\,J0717 is dominated by a bright relic, projected on the halo emission \citep{Rajpurohit2021b}. The halo in Abell 2142 is extended along the northwest to southeast direction \citep{Bruno2023}. To the north is a tailed radio galaxy (extending E-W). The radio beam size is indicated in the bottom left corner of each image. The contour levels are drawn at  $[1, 2, 4, 8 ...]\times 3\sigma_{\rm rms}$. Dashed contours depict the $-3\sigma_{\rm rms}$ contours. For MACS J0717+3745  image properties, see Table\,\ref{tab:imaging} (IM18 and IM19). The noise level of the LOFAR  image of Abell 2142 image is \ $\rm \sigma_{144\,MHz}=400\mu\,Jy\,beam^{-1}$.}
 \vspace{-0.3cm}
      \label{fig:MACSJ0717_A2142}
\end{figure*}

\setlength{\tabcolsep}{8pt}

\begin{table*}[!thbp]
    \caption{Clusters where the centrally located diffuse radio emission extent exceeds 2 Mpc}
    \centering
    \begin{tabular}{lcccccc c}
      \hline
      \hline  
        Cluster & \textit{Planck} name & $M_{500}$ & $z$ & $R_{500}$ & LLS & Classification&Reference \\
         & PSZ2 & ($10^{14} M_\odot$) & & (kpc) & (Mpc) & &  \\       
     \hline
        PLCK\,G287.0+32.9 & G286.98+32.90 & 14.69 & 0.390 & 1513 &3.4 &halo& This work  \\

        Abell 2744 & G266.04-21.25 & 9.84 & 0.306 & 1367 &2.6 &halo& This work \\
        Bullet & G266.04-21.25 & 13.10 & 0.296 & 1509 &2.4&halo &This work  \\
        
        MACS J0717+3745 & G180.25+21.03 & 11.49 & 0.549 & 1310 & 2.6&halo& This work \\    

     Abell 2142 & G044.20+48.66 & 8.77 & 0.089 & 1420 &2.5& halo& This work \\
        \hline
        Abell 2163 & G006.76+30.45 & 16.12 & 0.203 & 1676 & 2.4 &halo& \cite{Shweta2020} \\  
        Coma & G057.80+88.00 & 7.17 & 0.02 & 1357 & 2.0 & halo&\cite{Bonafede2022}\\    
        Abell 1758N & G107.10+65.32 & 7.99 & 0.280 & 1288 & 2.2 &halo&\cite{Botteon2018b}\\  
        Abell 2255 & G093.92+34.92 & 5.38 & 0.08 & 1210 & $\sim5$ &halo&\cite{Botteon2022} \\       
        \hline
        Abell 697 & G186.37+37.26 & 11.00 & 0.281 & 1432 &-& mega halo&\citet{Cuciti2022} \\      
        ZwlCl0634 & G167.67+17.63 & 6.65 & 0.174 & 1258 & -& mega halo&\citet{Cuciti2022} \\  
        Abell 665 & G149.75+34.68 & 8.86 & 0.182 & 1381 &-& mega halo&\citet{Cuciti2022}\\
        Abell 2218 & G097.72+38.12 & 6.59 & 0.171 & 1256 &-& mega halo& 
         \citet{Cuciti2022} \\     
       \hline 
    \end{tabular}
     \parbox{\textwidth}{\raggedright \small Notes: The LLS of mega halos are not known, however, by definition they are 2-3 Mpc in size. The LLS of the five radio halos presented in this work is measured at the $\geq 3\sigma_{\rm rms}$ level.} 
    \label{tab:large_halos}
\end{table*}

The new MeerKAT UHF and previously published uGMRT Band\,4 images of Abell 2744 are shown in Figure\,\ref{fig::A2744_halo}. Located at a redshift of $z=0.306$, this highly disturbed cluster is very rich in radio, X-ray, and optical wavelengths \citep{Govoni2001a,Owers2011, Merten2011, Kempner2004, Golovich2019, Jauzac2018, Venturi2013, Orru2007, George2017, Paul2019}. It is known to host a giant radio halo at its center, a 1.5~Mpc long relic to the northeast, and three additional fainter relics \citep{Pearce2017, Rajpurohit2021c, Knowles2022}. The LLS of the Abell 2744 halo is 2.6 Mpc at 675~MHz, exhibiting a roughly circular morphology. In our MeerKAT UHF map (0.55-1.0~GHz), we recovered the full 2.6 Mpc extent of the radio halo emission along with other known features (Figure\,\ref{fig::A2744_halo}). At 1.5 GHz, the halo extends to 2.1 Mpc, suggesting a steep spectrum in the outer regions \citep{Pearce2017}. By comparing the point-to-point radio brightness and spectral index with the X-ray brightness and temperature distribution, \cite{Rajpurohit2021c} found that the Abell 2744 halo consists of multiple components.

\begin{figure*}[!thbp]
    \centering
    \includegraphics[width=0.95\textwidth]{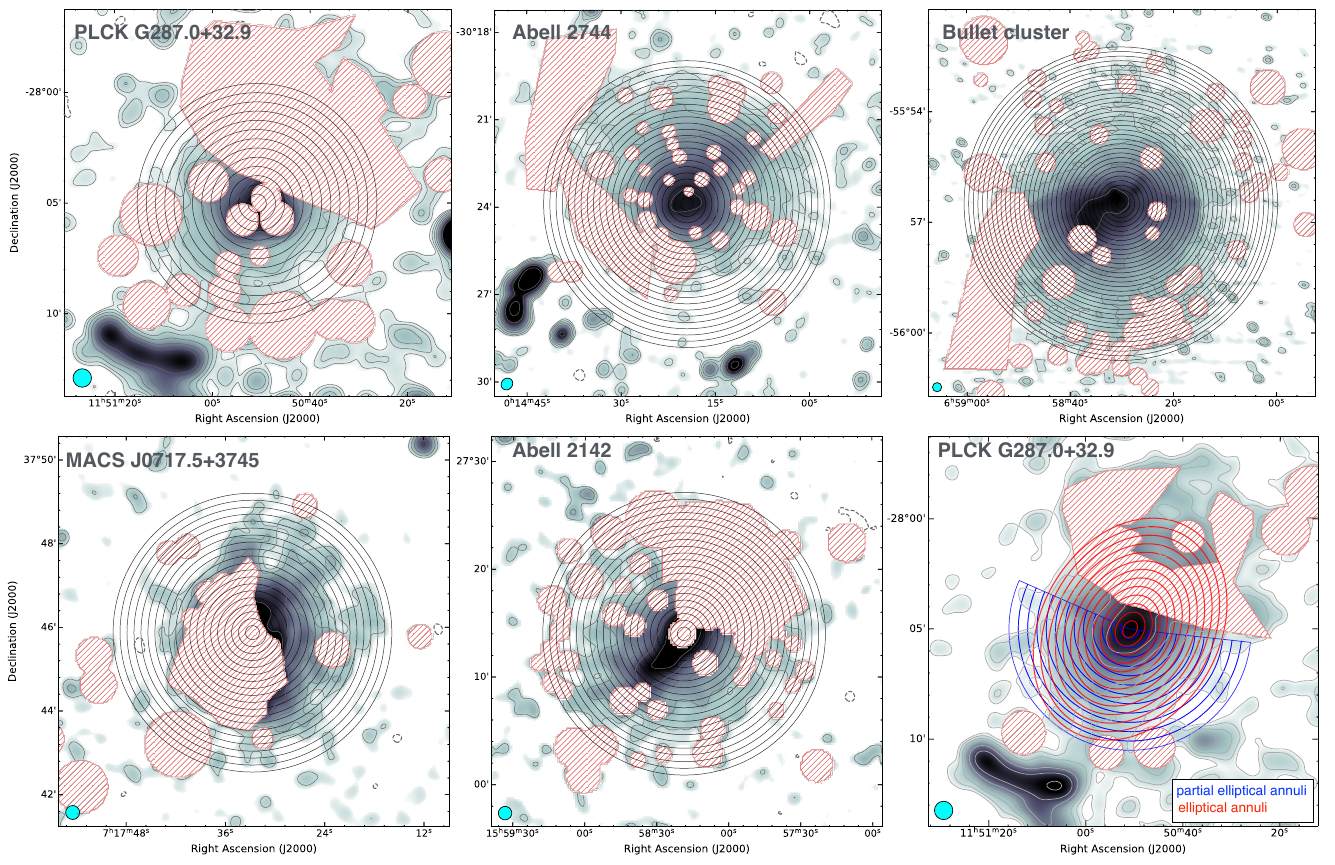}
 \caption{Concentric annuli used to extract the radial brightness profiles of the radio halos in PLCK\,G287.0+32.9 (50\arcsec\, resolution), Abell 2744 (25\arcsec\, resolution), Bullet (15\arcsec\, resolution), MACS\,J0717.5+3745 (20\arcsec\, resolution), and Abell 2142 \citep[75\arcsec\, resolution; image adopted from][]{Bruno2023}. The hatched areas indicate the masked regions. The width of each circular annulus is half the full width at half maximum of the beam size. The last panel illustrates the different azimuthal annuli (elliptical and partial elliptical) used to investigate the radial profiles of the PLCK\,G287+32.9 halo, with the resulting profiles shown in Figure\,\ref{different sectors} top right panel.  The last panel image of  PLCK\,G287.0+32.9 is obtained by subtracting discrete sources using an inner uvcut. For the rest of the image properties, see Table\,\ref {tab:imaging} (IM7, IM12, IM15, and IM19). We emphasize that the masked regions differ between the PLCK\,G287.0+32.9 maps with masking and those with uv-subtraction (see first and last panel images).}
      \label{fig:grid}
\end{figure*} 

Figure~\ref{fig:Bullet} shows the MeerKAT UHF and L-band images of the Bullet Cluster. The cluster is known to host a radio halo with LLS of about 2.4\,Mpc and a toothbrush-shaped relic to the east \citep{Sikhosana2023, Shimwell2014}. In our new UHF image, the full extent of the halo emission reported at L-band is recovered. The halo is asymmetric, extending primarily in the north-south and east-west directions. 

Figure~\ref{fig:MACSJ0717_A2142} shows the low-resolution LOw- Frequency ARray (LOFAR) High-Band Antenna (HBA) images of the halos in MACS J0717+3745 \citep[left panel:][]{Rajpurohit2021b}, and Abell 2142 \citep[right panel:][]{Bruno2023}, with LLS of about 2.6\,Mpc and 2.5\,Mpc, respectively. Similar to PLCK\,G287+32.9, both of these halos are asymmetrical, particularly MACS\,J0717+3745, which is known to host a bright filamentary relic at its center along with other linear filaments \citep{vanWeeren2017b, Rajpurohit2021a}. The LLS of the halos in MACS J0717+3745 and Abell 2142 is less than 2 Mpc at frequencies $> 650$~MHz \citep{vanWeeren2017b, Rajpurohit2021b,Riseley2024}. 

Clearly, in all cases, the radio emission is brightest at the halo core and decreases with radial distance. Radio halos with LLS$>$2~Mpc have been mainly detected at low frequencies, especially with observations $\leq 150$~MHz \citep[e.g.,][]{Botteon2018b, Shweta2020, Botteon2022, Bonafede2022, Rajpurohit2021b, Bruno2023}. Our results highlight the emergence of radio halos with large spatial extents even at high frequencies ($>$700 MHz) and demonstrate that their observed size is determined by the depth of the radio observations and uv-coverage, and their radio brightness. The detection of radio halos with large extents ($>$~2Mpc) indicates the presence of relativistic electrons and magnetic fields beyond $R_{500}$. It is, therefore, important to establish the observable quantities that allow a discrimination between these sources and others (e.g., mega halos).

\section{Radial properties of radio halos}

A handful of radio halos with remarkably large extents ($\geq2$~Mpc) are currently known, see Table\,\ref{tab:large_halos}. These systems are unique targets to test whether the physics of non-thermal components at such large scales is different from that at cluster-core scales and to probe the origin of the radio emission. The detection of halo emission, on such large scales, at multiple frequencies, in the five clusters analyzed in this work, provides a unique opportunity to investigate their properties and to shed light on the underlying particle acceleration mechanisms.

\subsection{Radial surface brightness profiles}

Assuming spherical symmetry, the surface brightness profile of radio halos is commonly described by fitting an exponential law of the form:
\begin{equation}			
I(r) = I_{0} e^{-r/r_e},
\end{equation}
where $I(r)$ is the surface brightness at radius $r$, $I_0$ is the central radio brightness, and $r_e$ is the e-folding radius. We emphasize that this model is not physically motivated. Although it has historically provided a reasonable description of halo profiles \citep[e.g.,][]{Orru2007, Murgia2009, Vacca2011, Cuciti2021}, the detection of substructures and multiple components in halos with new generation telescopes has highlighted its limitations \citep{Botteon2022a, Rajpurohit2021b, Botteon2023, Sikhosana2023}. Moreover, this model grossly oversimplifies the cluster morphology; merging systems can exhibit highly extended and irregular structures, for example, the PLCK G287.0+32.9 and Bullet halos. Nevertheless, this model is useful for comparative studies of different kinds of halos and, therefore, we adopt it in this paper to facilitate the comparison.  

For each cluster, we first computed the average radio brightness within concentric annuli centered on the peak of the halo emission, with the width of each annulus set to half the full width at half maximum of the beam size, see Table\,\ref{tab:fit_paramters}.  The circular annuli used are shown in Figure\,\ref{fig:grid}. The uncertainties in the profiles are estimated by accounting for systematic uncertainties due to flux scale/discrete sources subtraction (assumed to be 5-10\%) and statistical uncertainties associated with the rms level of an image. We assumed a flux density scale error of 10\% for LOFAR, uGMRT Band3 \citep{Shimwell2022, Chandra2017}, 8\% for uGMRT Band4 \citep{Chandra2017}, and 5\% for VLA and MeerKAT \citep{Perley2013}. We note that the flux density scale was verified by checking the spectra of compact sources in each field, we also refer to \cite{vanWeeren2017b, Pearce2017, Rajpurohit2021c, Rajpurohit2021b, Rajpurohit2021a} for details. We included only those annuli where the average surface brightness is $\geq 3\sigma_{\rm rms}$. The discrete, unrelated sources (namely, compact radio sources, radio galaxies, relics) were masked out from each map, as discussed in detail in Section\,\ref{sec::caveats}.

\begin{figure*}[!thbp]
    \centering
    \includegraphics[width=0.49\textwidth]{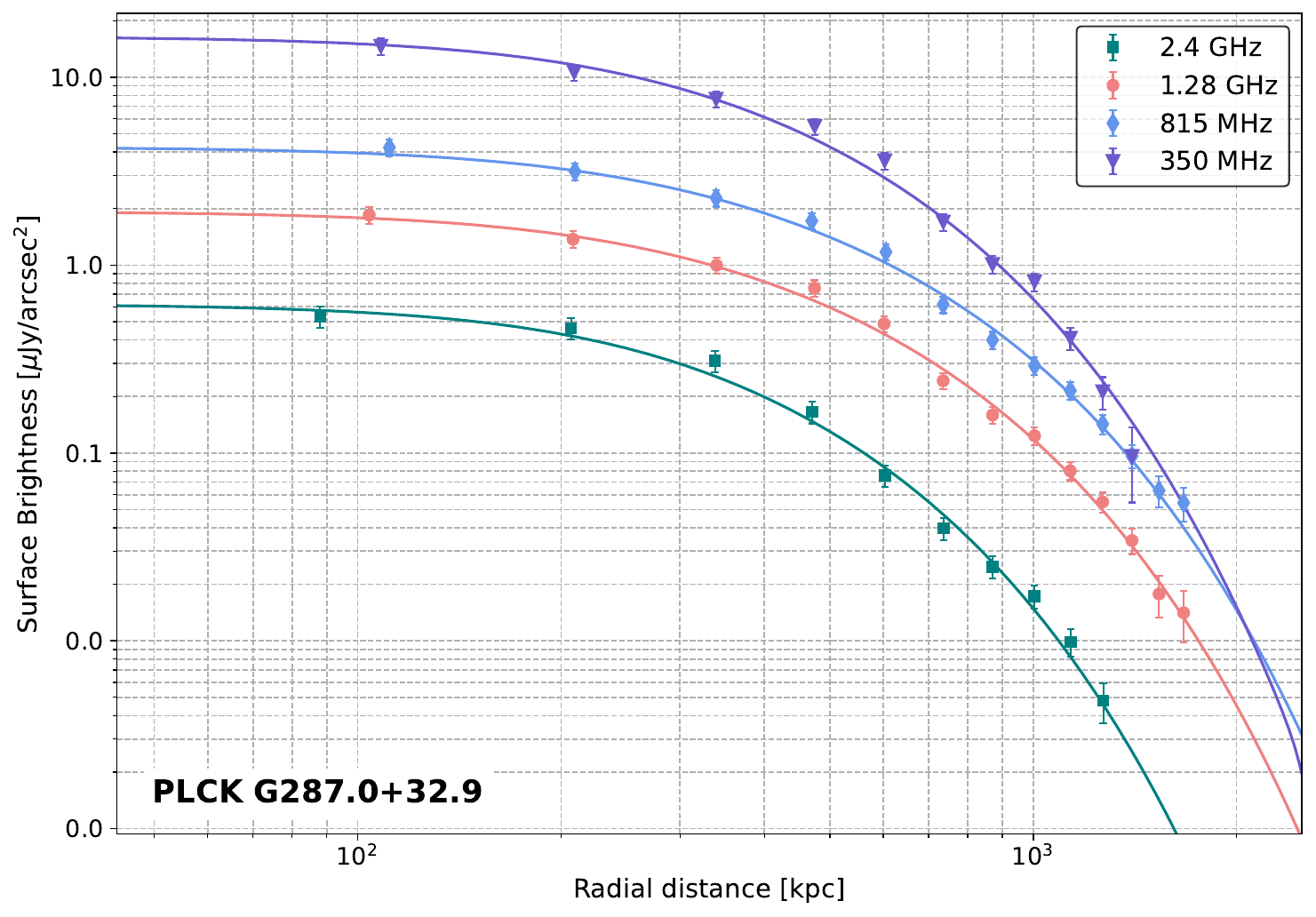}
    \includegraphics[width=0.49\textwidth]{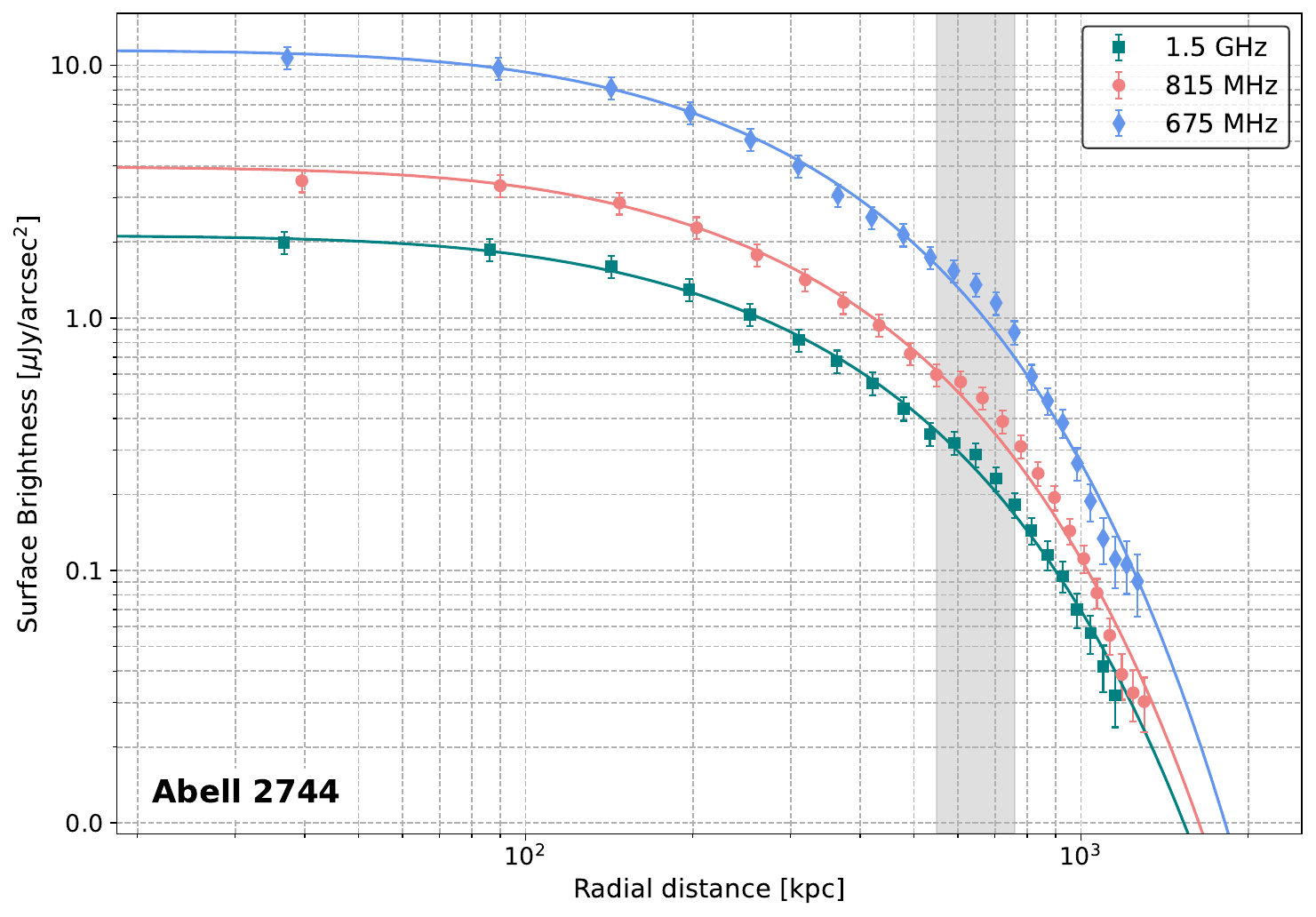}
    \includegraphics[width=0.49\textwidth]{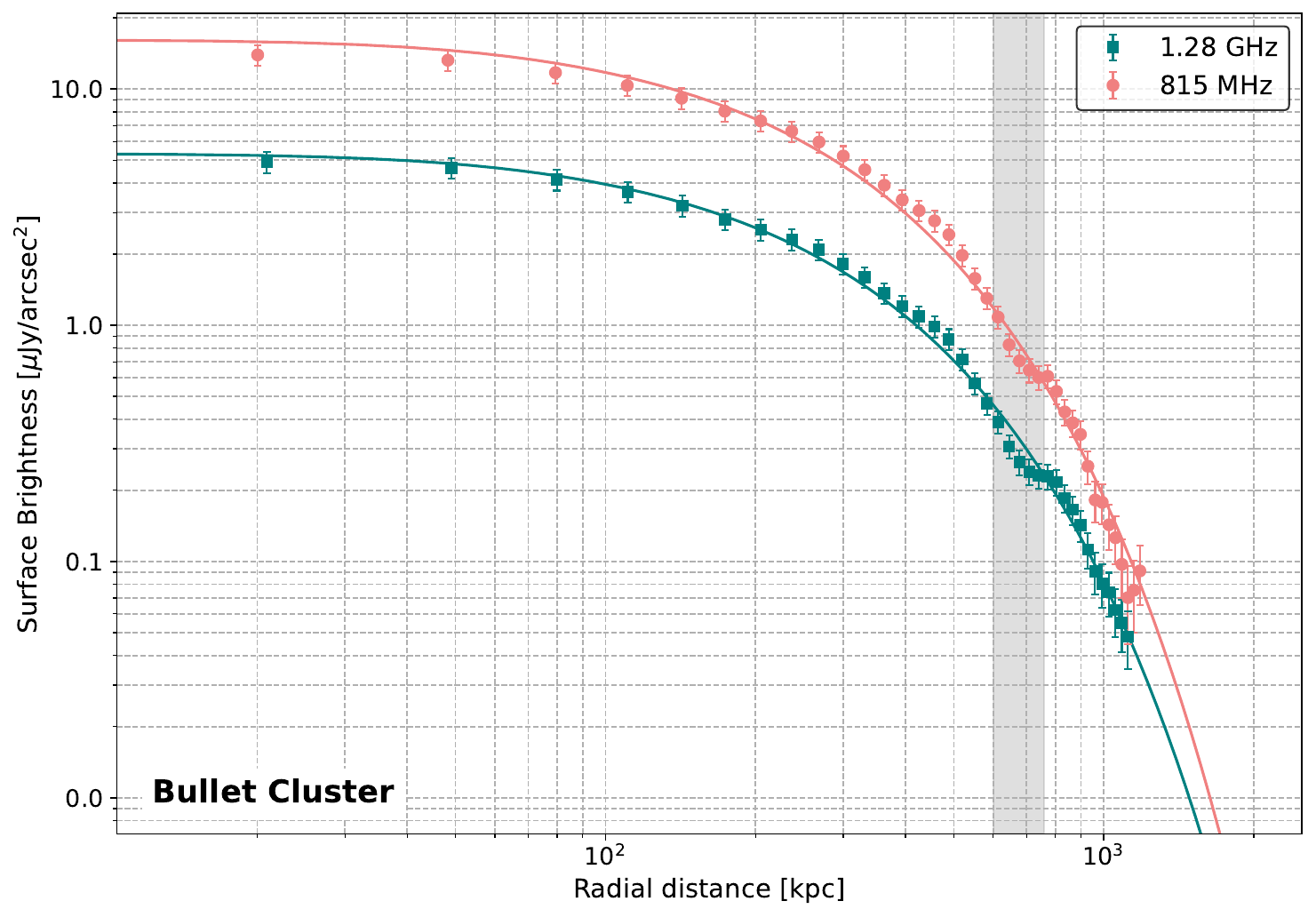}
    \includegraphics[width=0.49\textwidth]{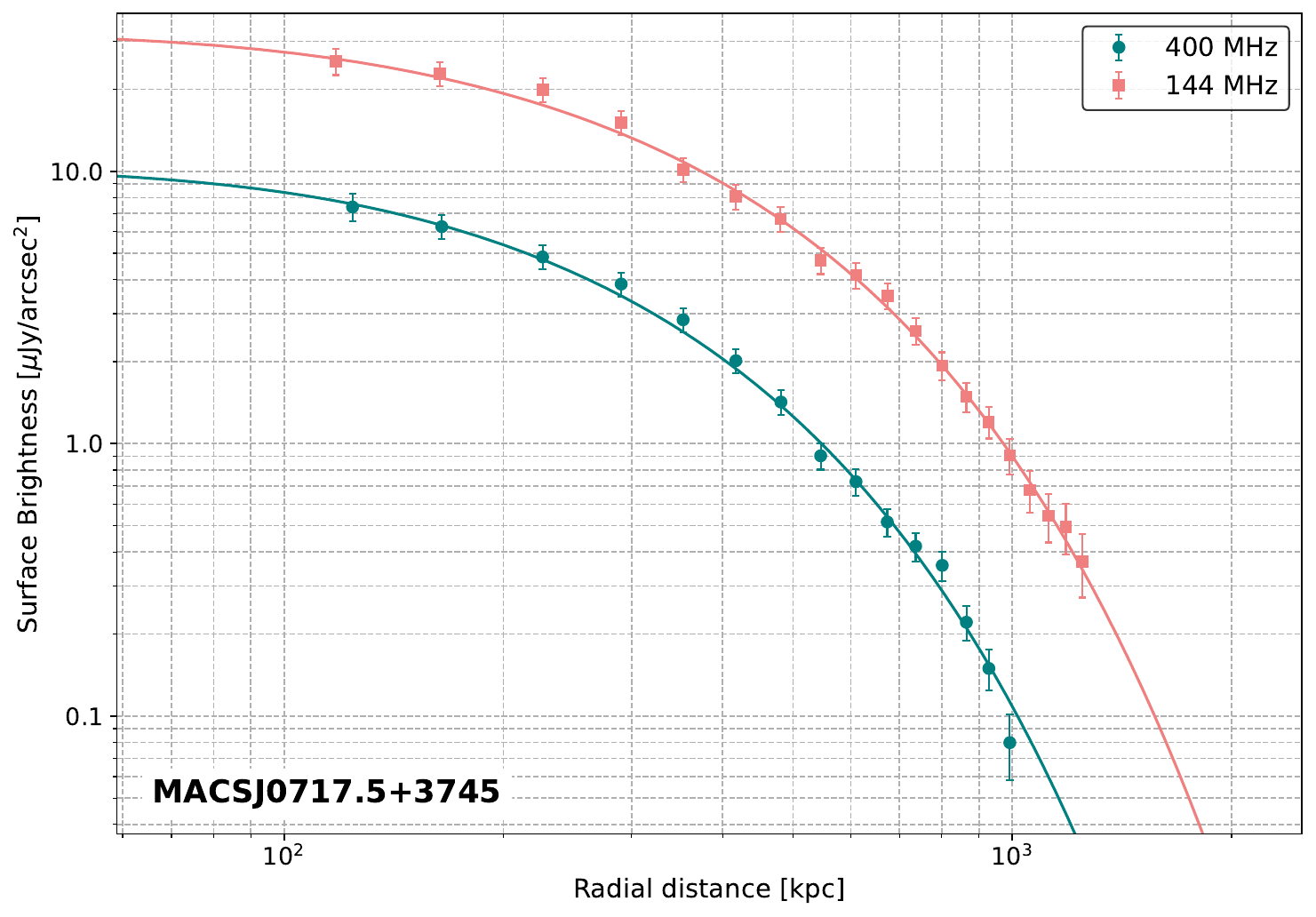}
 \caption{Radial radio profiles of the halos in PLCK\, G287.0+32.9, Abell 2744, Bullet and MACS J0717+3745 at difference frequencies.  The kinks in the radial profiles of the Bullet Cluster halo and Abell 2744 (gray shaded regions) are due to shock fronts. To avoid overlap, the surface brightness profiles for PLCK G287.0+32.9 at 2.4\,GHz, 815\,MHz, and 350\,MHz have been scaled by factors of 0.76, 1.5, and 2.5, respectively. Similarly, the Abell 2744 profile at 815\,MHz is scaled by a factor of 2. The surface brightness is measured in circular annuli (see Figure\,\ref{fig:grid} for the chosen annuli), and the data are fitted with single-component exponential profiles. The PLCK\, G287.0+32.9, Abell 2744, Bullet Cluster and MACS J0717+3745 profiles are extracted from 50\arcsec, 25\arcsec, 15\arcsec, and 20\arcsec\, resolution maps, respectively. The width of the annuli is half of the beam FWHM. At  all the observed frequencies, the halo emission can be described by a single component.}
      \label{fig:: PSZ_profiles}
\end{figure*} 

\begin{figure}[!thbp]
    \centering
    \includegraphics[width=0.49\textwidth]{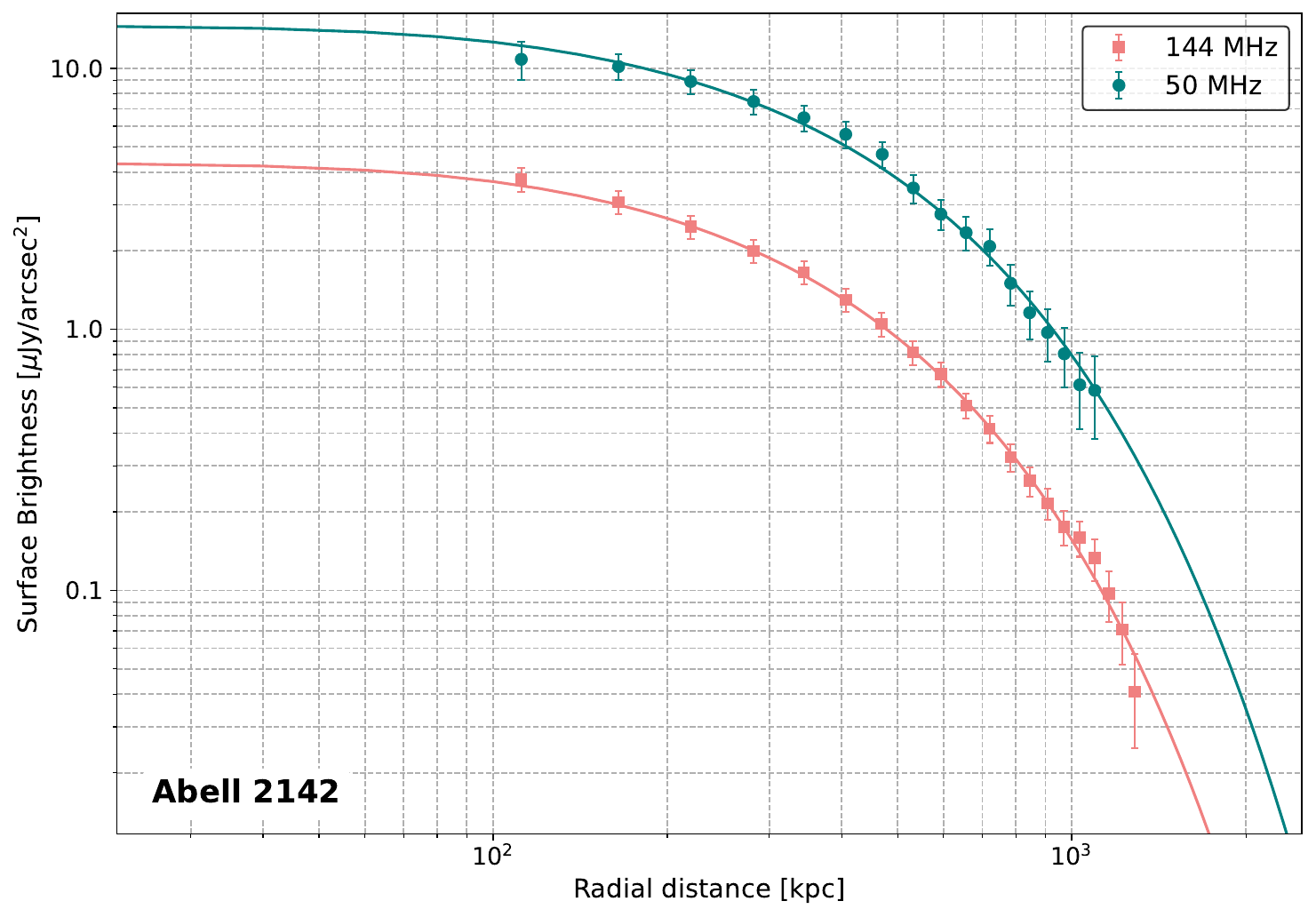}
 \caption{\textit{Left}: Abell 2142 halo profiles extracted from the 75\arcsec\, LOFAR HBA and LBA images. The average surface brightness is measured in circular annuli (see Figure\,\ref{fig:grid} for used annuli). The solid lines show the best-fitting single-component exponential profiles.}
      \label{fig::A2142_profile}
\end{figure}

In Figures\,\ref{fig:: PSZ_profiles} and \ref{fig::A2142_profile}, we show the resulting radial surface brightness profiles. For the first time, we present radial surface brightness profiles of radio halos out to $R_{500}$ as a function of frequency. Each point in these plots represents the mean radio brightness measured within each concentric circular annulus. The best fit parameters for each cluster are summarized in Table\,\ref{tab:fit_paramters}. Remarkably, in all cases, the halo emission can be described by a single exponential fit, despite the simplistic assumption of a symmetrical morphology and the highly disturbed dynamical nature of the clusters, which contain several substructures. 

The radial profiles of the PLCK\,G287.0+32.9 halo are well described by a single-component exponential fit at 350~MHz, 815~MHz, 1.28~GHz, and 2.4~GHz (see Figure\,\ref{fig:: PSZ_profiles} top-left panel). The best-fit e-folding radius varies with observing frequency, being smaller at lower frequencies. This is simply because the fainter outer regions are not detected at high signal-to-noise ratio at 350~MHz due to the lower sensitivity and missing short baselines of the uGMRT Band3 observations compared to the MeerKAT. However, between MeerKAT 815~MHz and 1.28~GHz, the total extent of the halo is similar and the corresponding best-fit e-folding radii indicate a clear radial steepening of the spectral index. 

Similarly, as shown in Figure\,\ref{fig:: PSZ_profiles} top-right panel, the radial profile of the halo in Abell 2744 can also be characterized by a single-component exponential fit at 675 MHz, 815~MHz and 1.5~GHz. It is worth noting that the profile exhibits a kink at radial distances between 600 and 800~kpc (shaded region). These annuli overlap with the southeastern relic (which is masked). We emphasize that while the halo in Abell 2744 shows multiple components in the radio versus X-ray analysis \citep{Rajpurohit2021c}, its radial profiles are consistent with a single component. 

The halo in the Bullet Cluster also shows a kink in its radial radio profile (Figure\,\ref{fig:: PSZ_profiles}, bottom-left panel), which coincides with the location of the bow shock detected via X-ray observations \citep{Markevitch2002}. This region is not masked as no relic-like emission is observed in the radio maps. Therefore, the kink is associated with the shock, similar to what is observed in Abell 2744, possibly suggesting a connection between shocks and turbulence. The radial profiles of both the Abell 2744 and Bullet Cluster halos show larger e-folding radii at higher frequencies (when comparing the same extent of the halo emission), hinting at relatively flatter spectral indices in the outermost regions. This is also evident from the 815~MHz profile, which steepens more rapidly at large radii than the 1.28~GHz profile. 

The 400\,MHz and 144\,MHz radial profiles of MACS\,J0717.5+3745 is shown in Figure\,\ref{fig:: PSZ_profiles}, bottom-right panel. Despite the complex morphology of the halo emission, the profiles are well described by a single exponential fit. The halo is more extended at lower frequencies, therefore, the e-folding radius is larger at 144\,MHz, suggesting the spectral steepening in the outermost regions. We note that the halo in MACS\,J0717.5+3745 also exhibits some filamentary features embedded in it \citep{vanWeeren2017b, Rajpurohit2021a}. However, those filaments are not associated with the halo emission \citep{Rajpurohit2022a}.

The radial profiles of the halo in Abell 2142 at 144 and 50~MHz are shown in Figure\,\ref{fig::A2142_profile}. A single-component exponential fit can describe radial profiles at both frequencies. Recently, \citet{Bruno2023} found that the same halo profile is better fit by a two-component exponential model. We used the same images for our analysis. The difference arises because we masked discrete sources based on high-sensitivity MeerKAT images \citep{Riseley2024}, which show a higher density of point sources, while \citet{Bruno2023} subtracted unrelated sources seen at 144~MHz from the uv-data (see Section\,\ref{sec::caveats} for details). We emphasize that although the Abell 2142 halo is known to consist of three components  based on a detailed radio and X-ray analysis \citep{Bruno2023}, its azimuthally averaged radial profile remains consistent with a single-component fit. We note that the inner mini-halo like component was masked in the present analysis due to the presence of compact sources in the central region.

\begin{figure}[!thbp]
    \centering
        \includegraphics[width=0.49\textwidth]{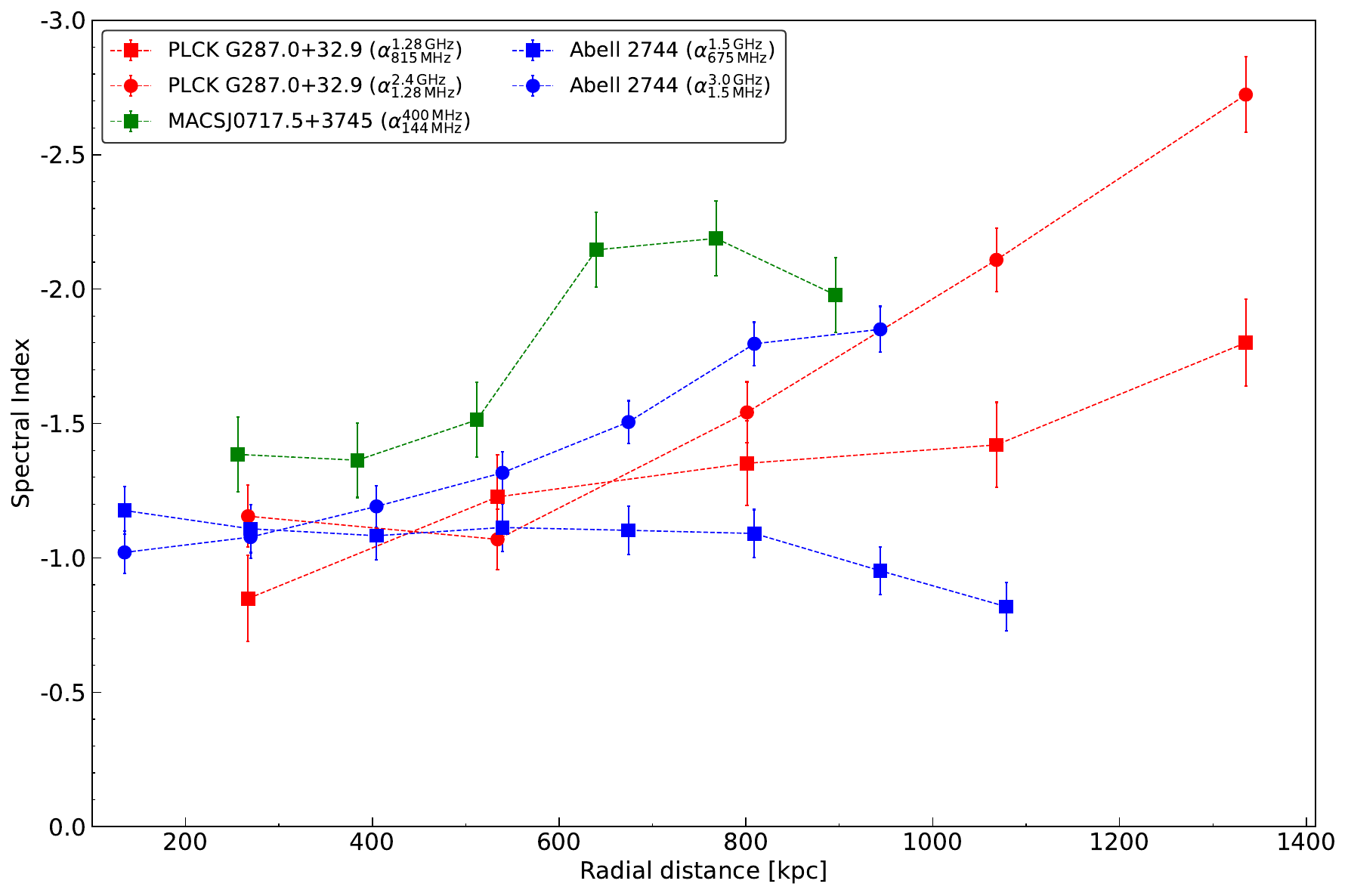}
 \caption{Radial spectral index profiles of PLCK\,G287.0+32.9, Abell 2744, and MACS\,J0717+3745. The error bars include flux scale uncertainties. To obtain these profiles, we measured average flux densities within concentric circular annuli from 50\arcsec\, (PLCK\,G287.0+32.9), 25\arcsec\, (Abell 2744), and 20\arcsec\, (MACS\,J0717+3745) resolution radio maps, with annulus widths of 50\arcsec, 30\arcsec, and 20\arcsec, respectively. All the observed spectral profiles of the halos show a radial steepening. }
      \label{fig::index_profiles}
\end{figure}

In summary, despite the complex morphologies of the radio emission,  the radial surface brightness profiles of the five halos with extents $>$2~Mpc, analyzed here, are consistent with other known halos ($<$2~Mpc). Thanks to the sensitivity of new generation telescopes, these five halos are detected at very large distances. Our sample consists of five radio halos but we do not find evidence of a shallower second component in the outer regions as reported by \citet{Cuciti2022} for four mega-halos.

\setlength{\tabcolsep}{10pt}
\begin{table*}[!thbp]
    \caption{Fitting results obtained from a single component exponential fit}
\begin{center}
    \begin{tabular}{lcccccc c}
      \hline
      \hline  
        Cluster & frequency & beam size & $\rm I_{0}$ & $\rm r_{e}$& $r_{\rm d}$& $\langle J_{\rm 1.4\,GHz} \rangle$ \\
         &MHz &  & $\rm \micro\,Jy \,arcsec^{-2}$& kpc & Mpc & $\mathrm{erg\,s^{-1}\,cm^{-3}\,Hz^{-1}}$\\       
     \hline
                            & 2400 & 50\arcsec & $1.43\pm0.10$ & $227\pm6$&1.3&\\
        PLCK\,G287.0+32.9   & 1280 & 50\arcsec & $2.60\pm0.19$ & $305\pm7$&1.7& $(2.70\pm0.21)\times10^{-42}$ \\
                            & 815 &  50\arcsec& $4.27\pm0.27$ & $325\pm8$&1.7&\\
                            & 350 &  50\arcsec& $10.87\pm0.72$ & $265\pm7$&1.5&\\
        \hline
       Abell 2744       & 1500 & 25\arcsec &$2.62\pm0.13$ & $271\pm6$&1.2& \\
                        & 815 & 25\arcsec &$4.99\pm0.22$ & $262\pm5$&1.3& $(2.25\pm0.12)\times10^{-42}$\\
                        & 675 & 25\arcsec &$7.26\pm0.33$ & $249\pm5$&1.3&\\
        \hline
        Bullet & 1280 & 15\arcsec&$6.18\pm0.23$ & $230\pm4$&1.2& $(5.99\pm0.25)\times10^{-42}$\\
                & 815 & 15\arcsec&$9.41\pm0.30$ & $217\pm3$&1.2&\\
        \hline
        MACS J0717+3745 & 400 & 20\arcsec&$14.4\pm0.9$ & $203\pm5$&1.0& $(3.7\pm0.29)\times10^{-42}$\\  
                    & 150 & 20\arcsec&$42.2\pm2.4$ & $259\pm6$&1.3&\\ 
               \hline
        Abell 2142  & 144&75\arcsec&$5.37\pm0.31$ & $284\pm8$&1.3&\\
                   & 50&75\arcsec&$17.9\pm1.3$ & $320\pm14$&1.1& $(1.43\pm0.09)\times10^{-43}$\\
                    
       \hline
    \end{tabular}
    \end{center}
        {Notes: Column 1: name of the cluster, Col. 2:  observed frequency, Col. 3: resolution of the input radio image, Col. 4:fitted central radio surface brightness, Col. 5:fitted e-folding radius, Col. 6: maximum detected radial distance, Col. 7: radio emissivity at 1.4~GHz}
        \vspace{0.5cm}
    \label{tab:fit_paramters}
\end{table*}

\subsection{Radial spectral index profile}

The radio observations presented in this study were obtained using different interferometers, each characterized by distinct uv-coverages. Consequently, careful consideration is required when comparing flux density measurements/extracting spectral index values for extended emission. To obtain the spectral index, we created maps using Briggs weighting with a robust parameter of $-0.5$ To ensure a consistent flux density distribution across all observed frequencies, we applied a common inner uv-cut at $200\lambda$, corresponding to the well-sampled shortest baseline of the uGMRT data, see see Table\,\ref{tab:imaging} for imaging properties (IM5-IM8, IM11-IM12, and IM16-IM19). For Abell 2142, we use the published 75\arcsec\ images. We note that the same images are used to estimate the radio power of halos.

We obtained radial spectral index profiles for the PLCK\,G287.0+32.9, Abell 2744, and MACS\,J0717.5+3745 halos by measuring average flux densities within concentric circular annuli from 50\arcsec, 25\arcsec, and 20\arcsec resolution radio maps, with annulus widths of 50\arcsec, 30\arcsec, and 20\arcsec, respectively. The chosen annulus widths ensure sufficient signal-to-noise to estimate the spectral index with small errors in the faint outer regions. For the spectral index profile of Abell 2142, we refer to \citet{Bruno2023}, while the profile for the Bullet Cluster halo will be presented in a follow-up paper.

The resulting azimuthally averaged spectral index profiles, centered on the point of peak surface brightness,  are shown in  Figure\,\ref{fig::index_profiles}. The halos in PLCK\,G287.0+32.9 and MACS\,J0717.5+3745 (previously reported by \citealt{Rajpurohit2021b}) show clear evidence of radial spectral steepening, also reported for other halos, e.g., Coma \citep{Bonafede2022}. The low frequency radial spectral index in the PLCK\,G287.0+32.9 halo steepens from $-0.9$ in the innermost region ($<250$\,kpc) to $-1.8$ at a distance of 1.3\,Mpc from the cluster center. The same trends are observed toward high frequencies with  steepening up to $-2.5$ in the outermost areas.  

For the halo in Abell 2744, the low frequency radial spectral index profile remains approximately constant (excluding the innermost region, $<$150 kpc) at around $-1.15$ to 800 kpc from the cluster center, consistent with the findings of \citet{Orru2007}. However, the high-frequency ($1.5-3$~GHz) spectral index profile shows a clear spectral steepening in the outer regions, also reported by \citet{Pearce2017}.  
Similarly, the MACS\,J0717.5+3745 halo also shows a radial steepening from $-1$ in the innermost region to $-1.8$ at 900\,kpc from the cluster center. We emphasize that in all cases, the outermost regions of the halo are steeper, as also reported for mega halos.    

For the halos in PLCK\,G287.0+32.9 and Abell 2744, the spectral index distribution is flatter at low frequencies and steepens toward higher frequencies, indicating clear spectral curvature. This is not surprising, as reported by \citet{Rajpurohit2021c, Rajpurohit2023}, the radio halos exhibit different spectral indices and curvature distributions. The radial spectral index steepening observed in these halos is consistent with the general expectations of the turbulent reacceleration models \citep{Brunetti2014}.

\begin{figure}[!thbp]
    \centering
    \includegraphics[width=0.48\textwidth]{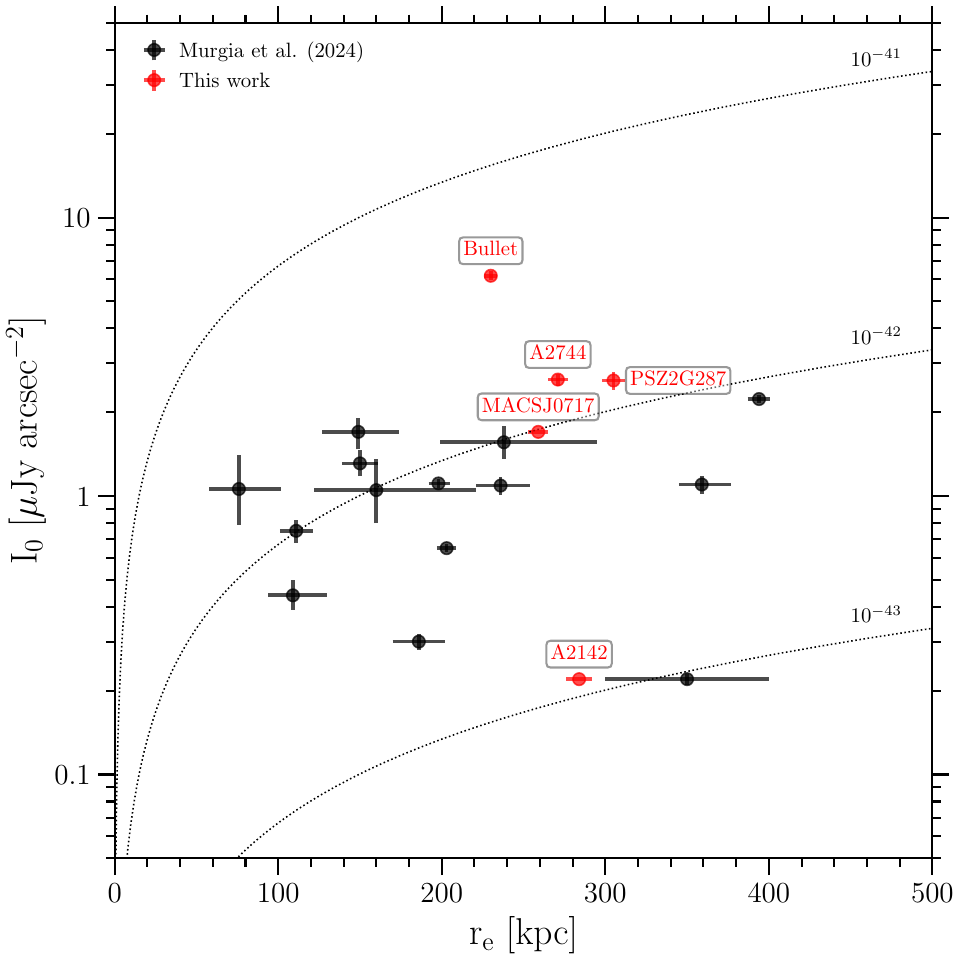}
 \caption{$I_0-r_{\rm e}$ plane at 1.4 GHz for the literature radio halos reported in \citet{Murgia2024} and those studied in this work. Dotted lines mark constant emissivities (in erg $s^{-1}$ cm$^{-3}$ Hz$^{-1}$ units) obtained using Eq.~\ref{eq:emissivity} assuming a redshift of $z=0.2$ and $\alpha=-1$ for reference.}
      \label{fig::emissivity_plot}
\end{figure}

\setlength{\tabcolsep}{10pt}
\begin{table*}[!thbp]
    \caption{Flux density and radio power of halos analyzed in this work}
\begin{center}
    \begin{tabular}{lcccccc c}
      \hline
      \hline  
        Cluster& $S_{\rm measured}$  & $S_{\rm fitted}$ & Measured radio power at 1.4~GHz (lower limit)  &  Fitted Radio power at 1.4~GHz\\
         &mJy & mJy & $\rm 10^{25}W\,Hz^{-1}$& $\rm 10^{25}W\,Hz^{-1}$&\\       
     \hline
     PLCK\,G287.0+32.9  & $23\pm2$& $42\pm0.5$ & $1.3\pm0.1$&$2.4\pm0.1$\\
        Abell 2744 &   $43\pm2$ & $47\pm4$&$1.3\pm0.1$ &$1.5\pm0.1$\\
           Bullet            &$94\pm12$ & $84\pm10$ & $2.7\pm0.4$&$2.3\pm0.3$\\
        MACS\,J0717    & $16\pm2$ & $14\pm02$ & $2.2\pm0.3$&$2.0\pm0.3$\\
          Abell 2142  & $26\pm2$ & $44\pm5$ & $0.05\pm0.01$&$0.08\pm0.01$\\                               
       \hline
    \end{tabular}
    \end{center}
       {Notes: Column 1: name of the cluster hosting the halo; Column  2:  Measured flux density of the halo at 1.5\,GHz for  Abell 2744 and MACS\,J0717 and 1.28\,GHz for PLCK\,G287.0+32.9, Bullet, and Abell 2142; Column 3: Fitted flux density of the halo, Col. 4: Radio power obtained from flux density reported in  Col 2; Column  Radio power obtained from flux density reported in Col 3.}
    \label{tab:power}
\end{table*}

\begin{figure*}[!thbp]
\centering
\includegraphics[width=0.505\textwidth]{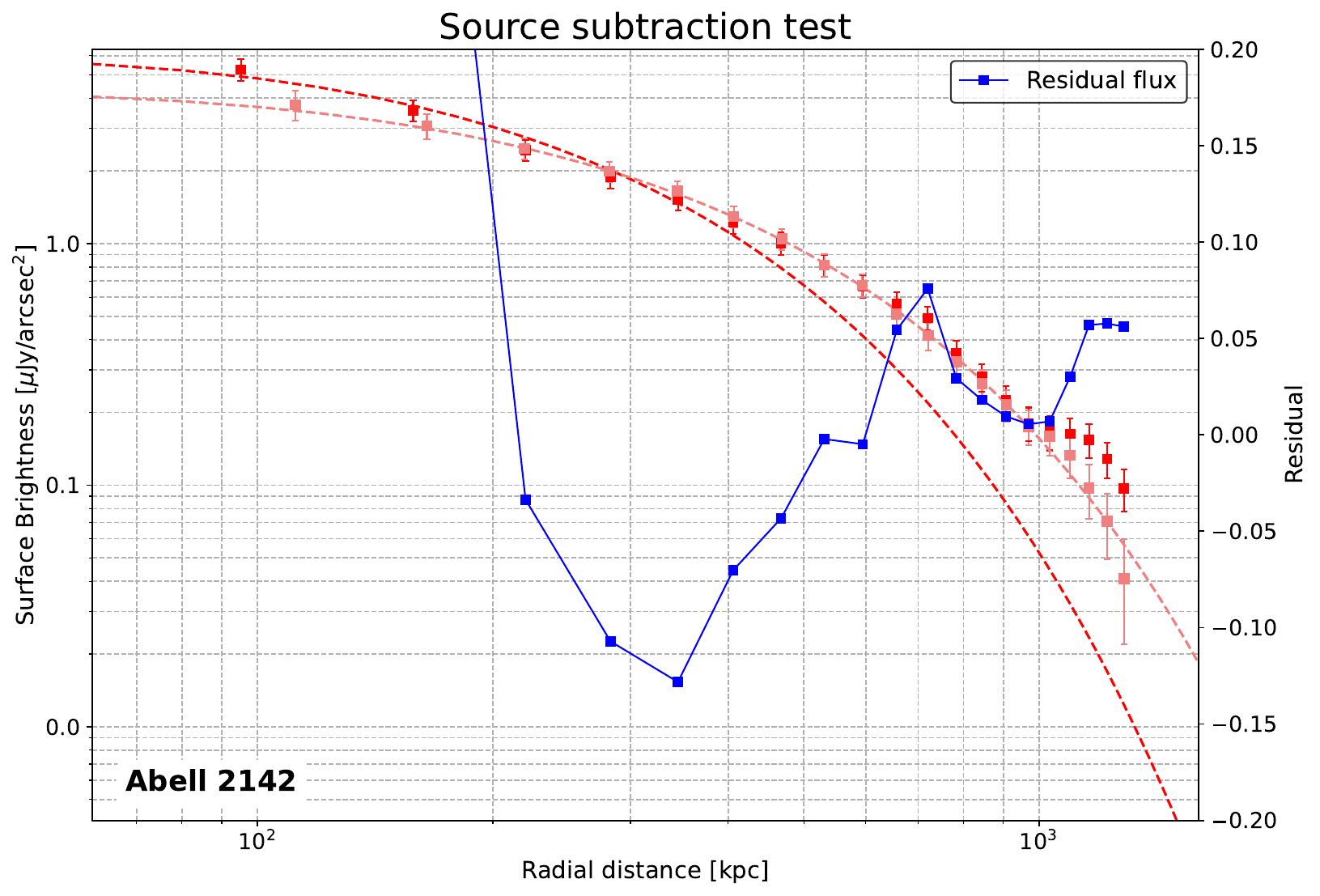}
 \includegraphics[width=0.49\textwidth]{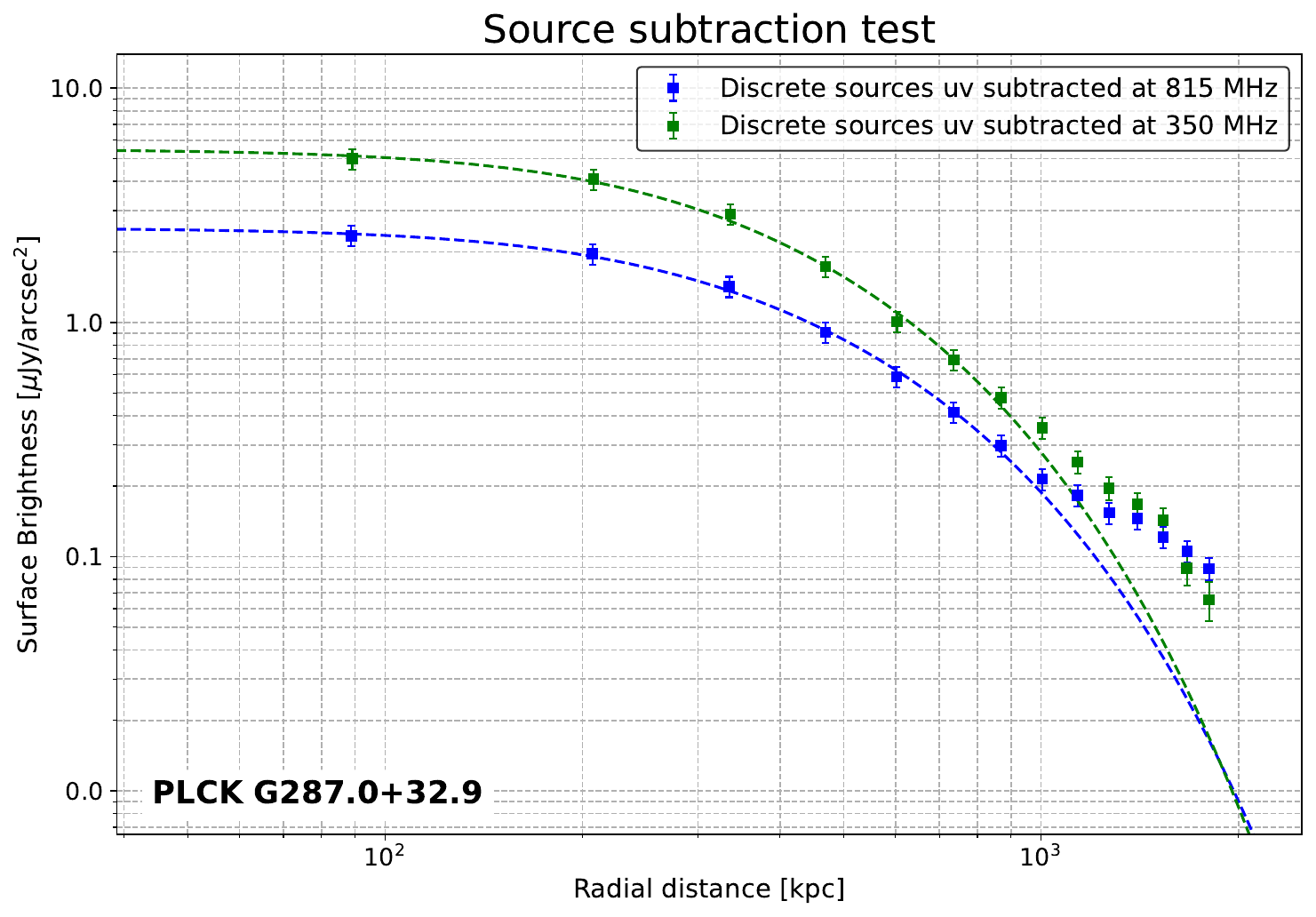}
\caption{Radial profiles of the halos in Abell 2142 (left) and PLCK\,G287.0+32.9 (right). Profiles fitted with a single-component exponential model are shown with dashed lines. The used annuli are shown in Figure\,\ref{fig:grid} \textit{Left}: Radial profile of the Abell 2142 halo at 144 MHz, extracted from an image where unrelated sources were masked based on the LOFAR HBA map (red data points) and from an image where unrelated sources were masked using the sensitive 1.3 GHz MeerKAT map (lightcoral data points). The blue data points/line show the residual radial profile obtained by subtracting the $uv$-subtracted sources profile from the masked out sources profile. \textit{Top right}: PLCK\,G287.0+32.9 halo profiles at 815\,MHz and 350~MHz obtained after subtracting discrete sources from the $uv$-data. Both plots exhibit a second outer component resulting from the incomplete subtraction of discrete sources.}
\label{different sectors}
\end{figure*}

\subsection{Radio emissivity and radio power}

Radio halos are observed to exhibit a relatively narrow distribution of radio emissivity, with a characteristic value of approximately $\rm 10^{-42}\,erg s^{-1}\,cm^{-3}\,Hz^{-1}$ \citep[e.g.,][]{Murgia2009}. This is based on the assumption of a homogeneous/filling $\rm factor=1$ emitting volume. The emissivity of mega-halos (second component) is reported to be approximately $10^{-44}\,\rm erg s^{-1}\,cm^{-3}\,Hz^{-1}$, about a factor of 20-25 lower than that of classical radio halos \citep{Cuciti2022}. 

To estimate the volume-averaged radio emissivity ($\langle J \rangle$) of the halos in this work, we adopt the formalism presented in \citet{Murgia2009}. Assuming a spherical geometry for the radio halos, the emissivity is computed using:

\begin{equation}\label{eq:emissivity}
\langle J \rangle \simeq 7.7 \times 10^{-41} (1 + z)^{3 - \alpha} \cdot \frac{I_0}{r_e} \,\,\, \left( \mathrm{erg\,s^{-1}\,cm^{-3}\,Hz^{-1}} \right),
\end{equation}

where $I_0$ is the central surface brightness in $\mu$Jy/arcsec$^2$, $r_e$ is the e-folding radius in kpc, and $\alpha$ is the integrated radio spectral index. The term $(1 + z)^{3 - \alpha}$ takes account of both the $k$-correction and cosmological dimming of the surface brightness with redshift. The values of $I_0$ and $r_e$ are obtained from fitting an exponential model to the radial surface brightness profiles. The radio emissivities for the halos in our sample are reported in Table\,\ref{tab:fit_paramters}. 

We find that the observed halos display a comparable emissivity of approximately $10^{-42}\,\mathrm{erg\,s^{-1}\,cm^{-3}\,Hz^{-1}}$. In Figure~\ref{fig::emissivity_plot}, we present the $I_0 - r_e$ plane at 1.4\,GHz for our sample (red points), along with a comparison to a sample of 14 known halos from \citep{Murgia2009, Vacca2011, Murgia2024}. We note that \cite{Murgia2024} also masked unrelated sources, making the results directly comparable. Despite their different LLS, we find that the emissivities of halos with LLS$>2$~Mpc are remarkably similar to those of classical halos \citep{Murgia2024}. Moreover, the e-folding radius of these five halos is only about 50~kpc larger than the average e-folding radius reported by \citet{Murgia2024}

The halo in Abell 2142 seems different as its emissivity is about a factor of 10 lower than the other halos with LLS$>2$~Mpc, namely $\langle J \rangle=( 1.43\pm0.09) \times 10^{-43}\,\mathrm{erg\,s^{-1}\,cm^{-3}\,Hz^{-1}}$ (estimated using an integrated spectral index of $-1.2$, i.e., mean spectral index of the three components). The prototype radio halo in Coma has the lowest emissivity observed so far at 1.4~GHz, with $\langle J \rangle \simeq 5 \times 10^{-44}\,\mathrm{erg\,s^{-1}\,cm^{-3}\,Hz^{-1}}$ \citep{Murgia2024}. We note that the Coma and Abell 2142 halos are peculiar, although both are nearby clusters and their radii are comparable to other halos with large extents, their central brightnesses are about an order of magnitude fainter (see Figure~\ref{fig::emissivity_plot}). The lower emissivities of the Abell 2142 and Coma halos suggests that halos with such low emissivity may not be uncommon. 

We estimated the monochromatic radio power for each halo as follows:
\begin{equation}
P_{\nu} = 4\pi D_L^2\, S_{\nu}\, (1 + z)^{-(1 + \alpha)},
\end{equation}
where $D_L$ is the luminosity distance to the cluster, $S_{\nu}$ is the flux density at a frequency $\nu=1.4$\,GHz, and  $\alpha$ is the spectral index used in the $k$-correction. For the measured flux density of these halos from 1.5 or 1.38\,GHz VLA or MeerKAT maps, see Table\,\ref{tab:power}.  We emphasize that flux densities were measured within the halo region where the signal exceeds the $\geq 3\sigma_{\rm rms}$. The derived radio powers, at 1.4~GHz, are $P_{\rm PLCK\,G287}=(1.3\pm0.1)\times10^{25}\,\mathrm{W\,Hz^{-1}}$ (using $\alpha=-1.15$), $P_{A2744}=(1.4\pm0.1)\times10^{25}\,,\mathrm{W\,Hz^{-1}}$ \citep[using $\alpha=-1.15$;][]{Rajpurohit2021c,Pearce2017}, $P_{\rm MACS\,J0717}=(2.2\pm0.3)\times10^{25}\,\mathrm{W\,Hz^{-1}}$ \citep[using $\alpha=-1.5$;][]{Rajpurohit2021b}, $P_{\rm Bullet}=(2.7\pm0.4)\times10^{25}\,\mathrm{W\,Hz^{-1}}$ (using $\alpha=-1.1$), $P_{A2142}=(5.2\pm1.3)\times10^{23}\,\mathrm{W\,Hz^{-1}}$ \citep[using $\alpha=-1.2$;][]{Riseley2024}. These values represent lower limits, as part of the halo emission is excluded/masked due to the presence of discrete sources.

\begin{figure}[!thbp]
    \centering
    \includegraphics[width=0.49\textwidth]{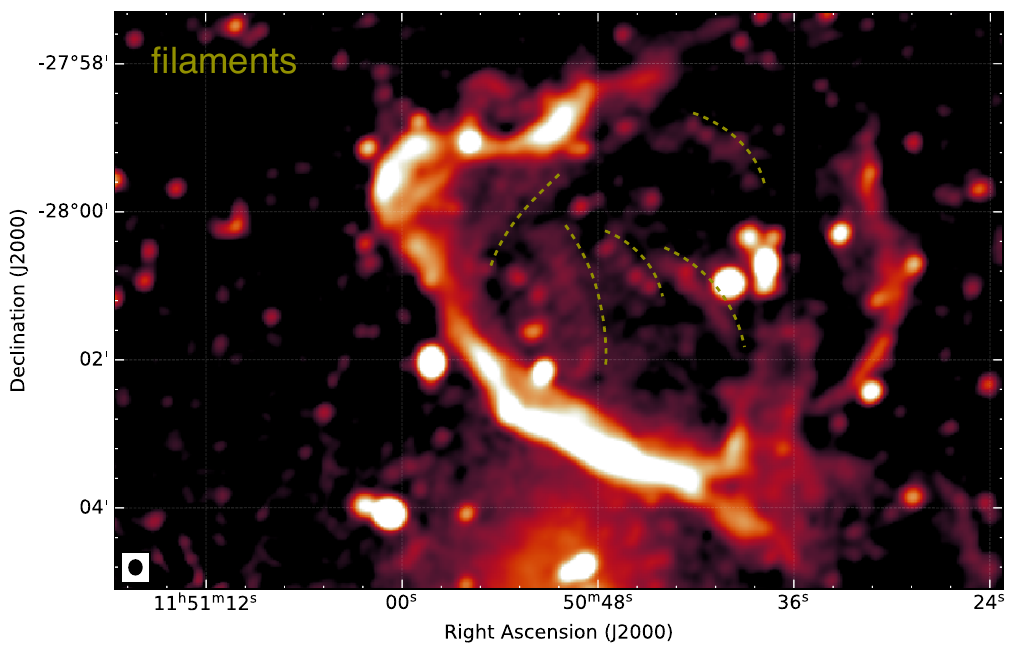}
       \includegraphics[width=0.49\textwidth]{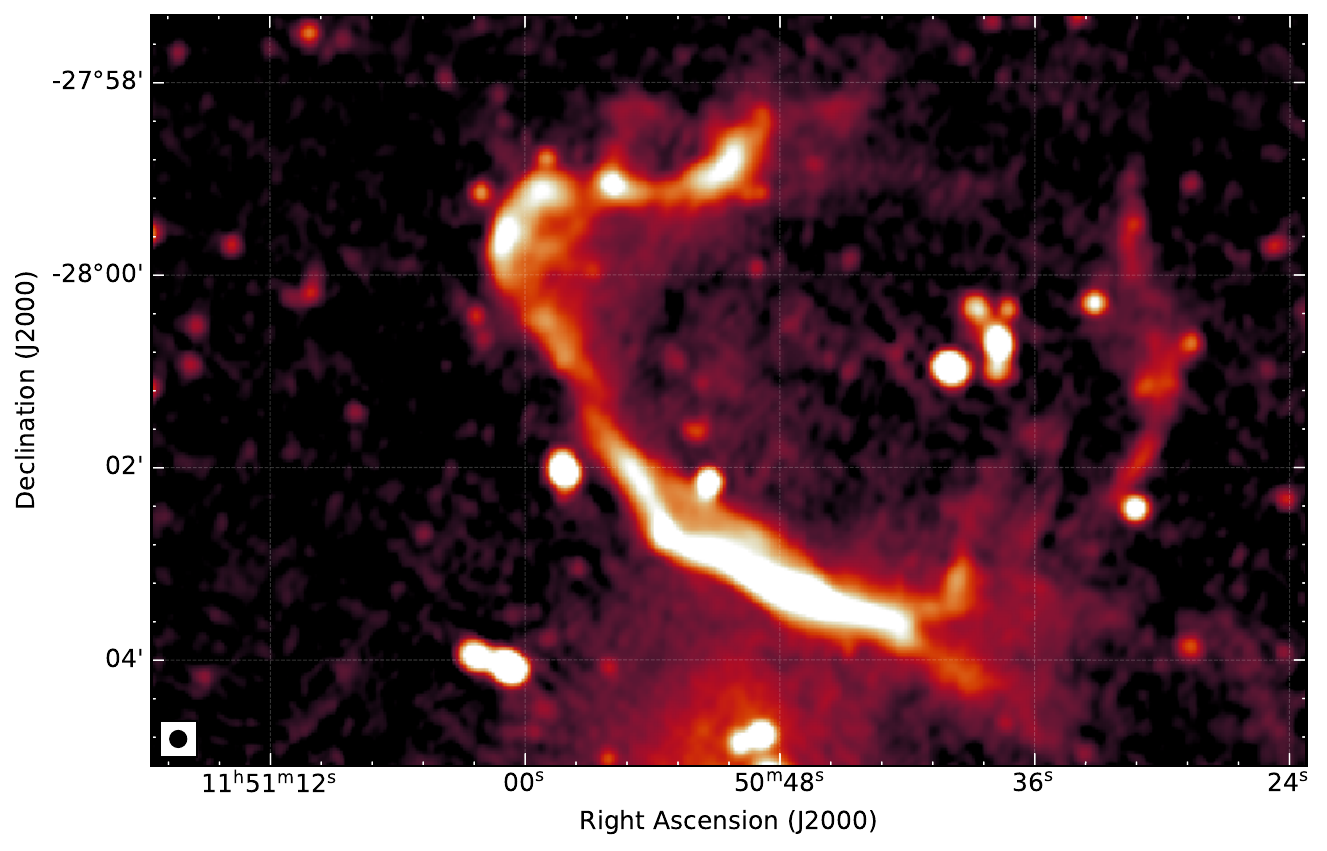} 
 \caption{Zoom-in view of the MeerKAT UHF (top) and uGMRT Band~3 (bottom) 10\arcsec resolution images, revealing substructures (marked with green curves) embedded in the northern part of the halo (Rajpurohit et al. in prep). These features are identified in sensitive high frequency observations but not detected in the less sensitive low-frequency images. When not masked, these features result in a second component at large radii in the radial profiles. }
      \label{fig::subfeatures}
\end{figure}

\begin{figure*}[!thbp]
    \centering
    \includegraphics[width=0.9\textwidth]{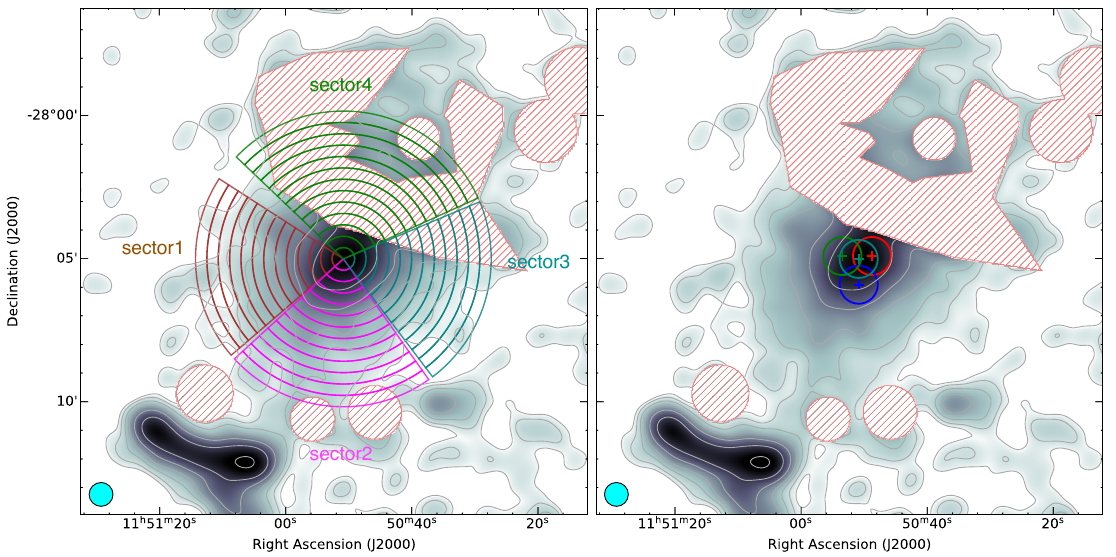}
 \caption{\textit{Left}: Sectors used to extract the PLCK\,G287.0+32.9 halo radial profiles shown in Figure\, \ref{different sectors1} middle panels. \textit{Right}: Different annuli centers (marked with "+" symbol and a circle of radius 10\arcsec) used to extract radial profiles shown in Figure\, \ref{different sectors1}, bottom panels.  These sectors are overlaid on the 50\arcsec MeerKAT UHF uv-subtracted point source image.}
      \label{fig::regions}
\end{figure*}

Following \citet{Murgia2009}, we also estimated the flux density of the halos at 1.4~GHz by integrating the surface brightness of the best-fit exponential model out to a radius of $r= 3r_e$:
\begin{equation}
S= 2\pi f \cdot r_e^2 I_0 \quad \text{(mJy)},
\end{equation}
where $r_e$ and $I_0$ are in units of arcseconds and $\rm mJy\,arcsec^{-2}$, respectively, and $f\approx0.8$ accounts for the enclosed flux within $3r_e$. The fitted flux densities and the corresponding radio powers are summarized in Table~\ref{tab:power}. The fitted and measured flux densities of the radio halos in Abell 2744, MACS J0717+3745, and the Bullet Cluster are consistent. However, the fitted flux densities for the halos in PLCK G287.0+32.9 and Abell 2142 are lower than the measured values by a factor of two. \citet{Murgia2009} reported that a tight relation exists between the total radio power of halos and e-folding radius, i.e., $P_{1.4} \propto r_e^3$. The halos analyzed in this work are consistent with this relation and align well with known classical halos. Moreover, powerful radio halos are expected to be more extended, corresponding to larger e-folding radii  \citep{Cassano2007, Murgia2009}.  This also implies that powerful halos should be detectable out to larger radii.

The lower limits and the fitted radio power imply that, excluding the halo in Abell 2142, the remaining systems lie above the established scaling relation between cluster mass and radio power observed for halos at 1.4~GHz. In contrast, the halo in Abell 2142 lies below this relation, as reported by \cite{Riseley2024, Bruno2023}. This also implies that radio powers of halos reported in the literature may be either underestimated \citep{Shimwell2014, Bonafede2014} or overestimated \citep{vanWeeren2009, Bonafede2009a, venturi2017} due to poor sensitivity and resolution. This in turn impacts the established $P_{\rm 1.4\,GHz}-M_{500}$ scaling relations of radio halos.

\section{Caveats on the profile extraction}
\label{sec::caveats}

The extraction of radial profiles is subject to systematic factors that can lead to misinterpretation of the results. In this section, we discuss key considerations that impact the interpretation of the halo radial profiles, including the subtraction/masking of discrete sources, the choice of sectors, and the selection of the annulus/sector center. 

\subsection{Source subtraction/masking}
To assess the impact of removing unrelated radio sources from halo emission, we created profiles using two methods: (1) masking out all discrete sources (2) subtracting them from uv-data using a uv-cut. The second method is a commonly used approach in the literature. 

In the first method, discrete sources (compact) were masked based on the high and low frequency radio maps. We emphasize that to mask out extended unrelated sources, such as radio galaxies or relics embedded within the halo, we always use the lowest frequency maps because these sources are typically more extended at low frequencies.  We find that masking discrete sources using high and low frequency maps results in significantly different radial profiles. For example, in Figure\,\ref{different sectors} left panel, we show the radial profiles of the Abell 2142 halo obtained by masking sources based on the LOFAR 144\,MHz and MeerKAT 1.3\,GHz. Similar trends are observed in the radial profile of the other four halos analyzed. As evident when unrelated sources are masked out based on the LOFAR HBA map (where the number density of compact sources is low), we see the presence of a second component at larger radius, see Figures\,\ref{different sectors} left. In contrast, the profile obtained by masking out unrelated sources based on the sensitive high frequency map can be described by a single-component and shows no evidence of any second component. 

The secondary component is not real and results from the incomplete masking of unrelated sources. In the inner region, the bright halo emission dominates. Therefore, incomplete source subtraction does not significantly impact the profiles. However, at larger radii, where the diffuse emission is weaker, the contribution from unmasked discrete sources becomes prominent relative to the halo emission, introducing an apparent shallower secondary component. This is shown in Figure\,\ref{different sectors} left panel, where we present the residual profile (blue line) of the Abell 2142 halo, obtained by subtracting the radio profile with unrelated sources masked using the low-frequency map from the profile where masking was performed using the high-frequency map. As evident, in the outer regions ($>$600~kpc), where the halo surface brightness is low, the contribution from unmasked discrete sources becomes significant, resulting in an apparent second component at larger radial distances. Therefore, high frequency radio maps should be preferred for masking discrete sources as they typically show a higher density of point sources than low frequency maps due to better sensitivity.

In the second method, we subtracted discrete (compact and/or extended) sources from the uv-data. In this method, we first created an image using an inner uv-cut of $3.8\,k\lambda$ (with robust=$-0.5$), which was subsequently used as a model and subtracted from the uv-data. We emphasize that the same uv-cut was applied to all sources, as this cut effectively recovered the flux from discrete sources. In none of the resulting maps was halo emission seen in the model image. The data were then re-imaged without any uv-cut.  In Figure\,\ref{fig:grid} last panel, we show the MeerKAT UHF (815\,MHz) uv-subracted point source image. The resulting radial profiles for the PLCK\,G287.0+32.9 halo at 815~MHz and uGMRT 350~MHz is shown in Figure\,\ref{different sectors} right panel. The regions covering the northern relic and the tailed radio galaxy to the northeast were manually masked. We emphasize that in this case, the northernmost part of the halo is not masked and included in the radial profiles. Unlike the profile obtained by directly masking out discrete sources from the image (Figure\,\ref{fig:: PSZ_profiles} top left), this profile shows a two-component structure at both 815~MHz and 350~MHz. However, the second component is introduced due to the incomplete subtraction of discrete sources.

\begin{figure*}[!thbp]
\centering
\includegraphics[width=0.49\textwidth]{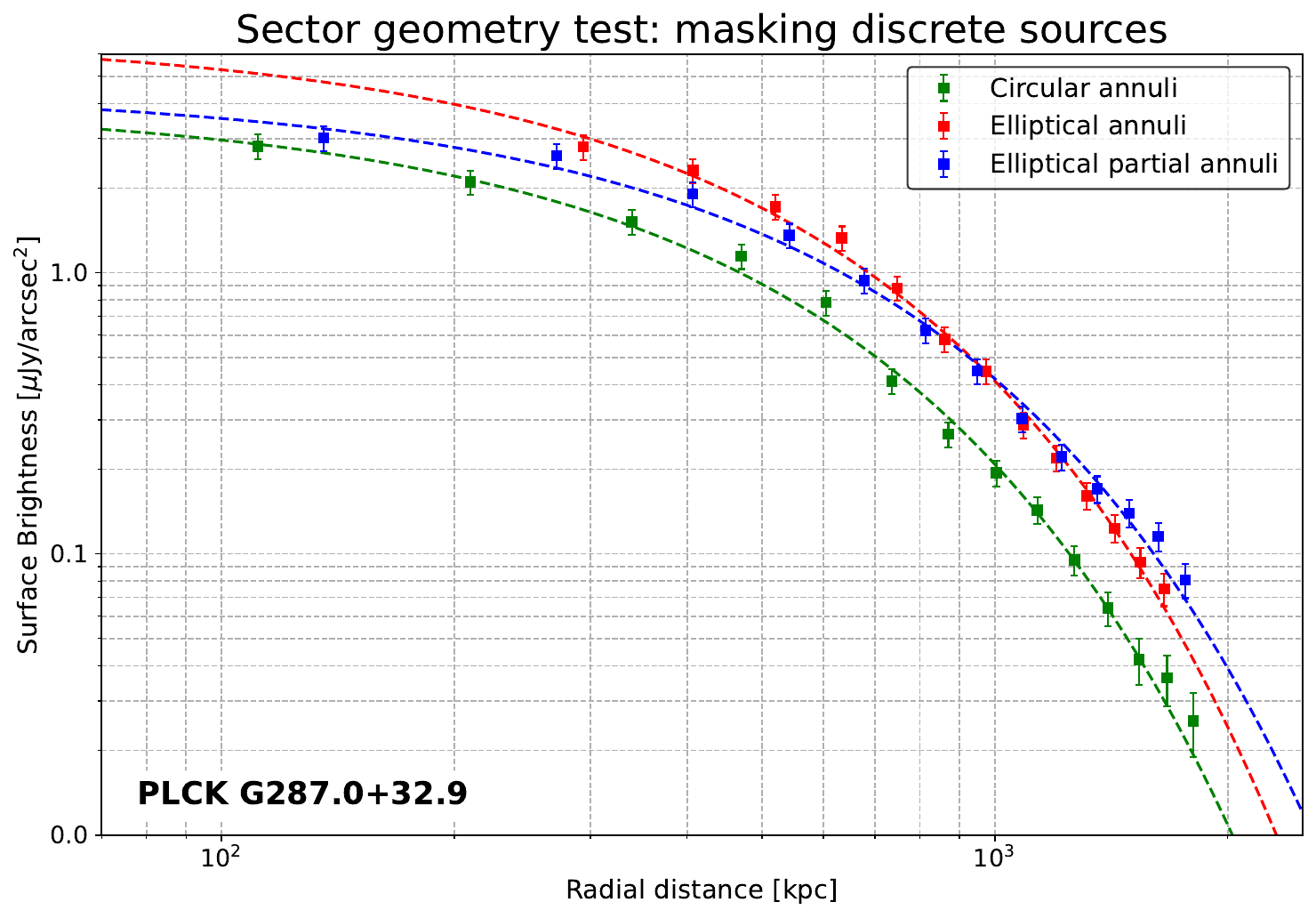}
\includegraphics[width=0.49\textwidth]{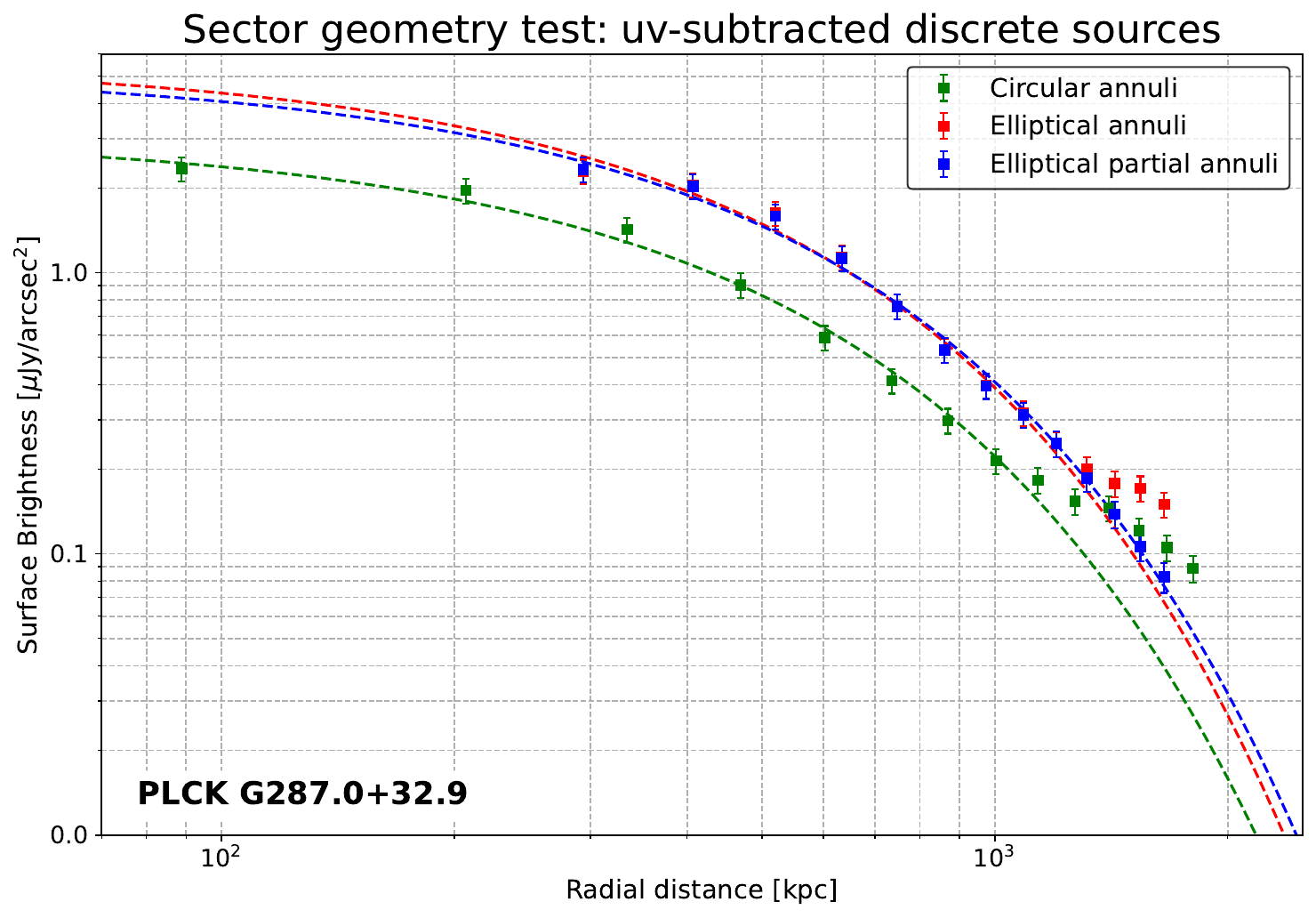} 
\includegraphics[width=0.49\textwidth]{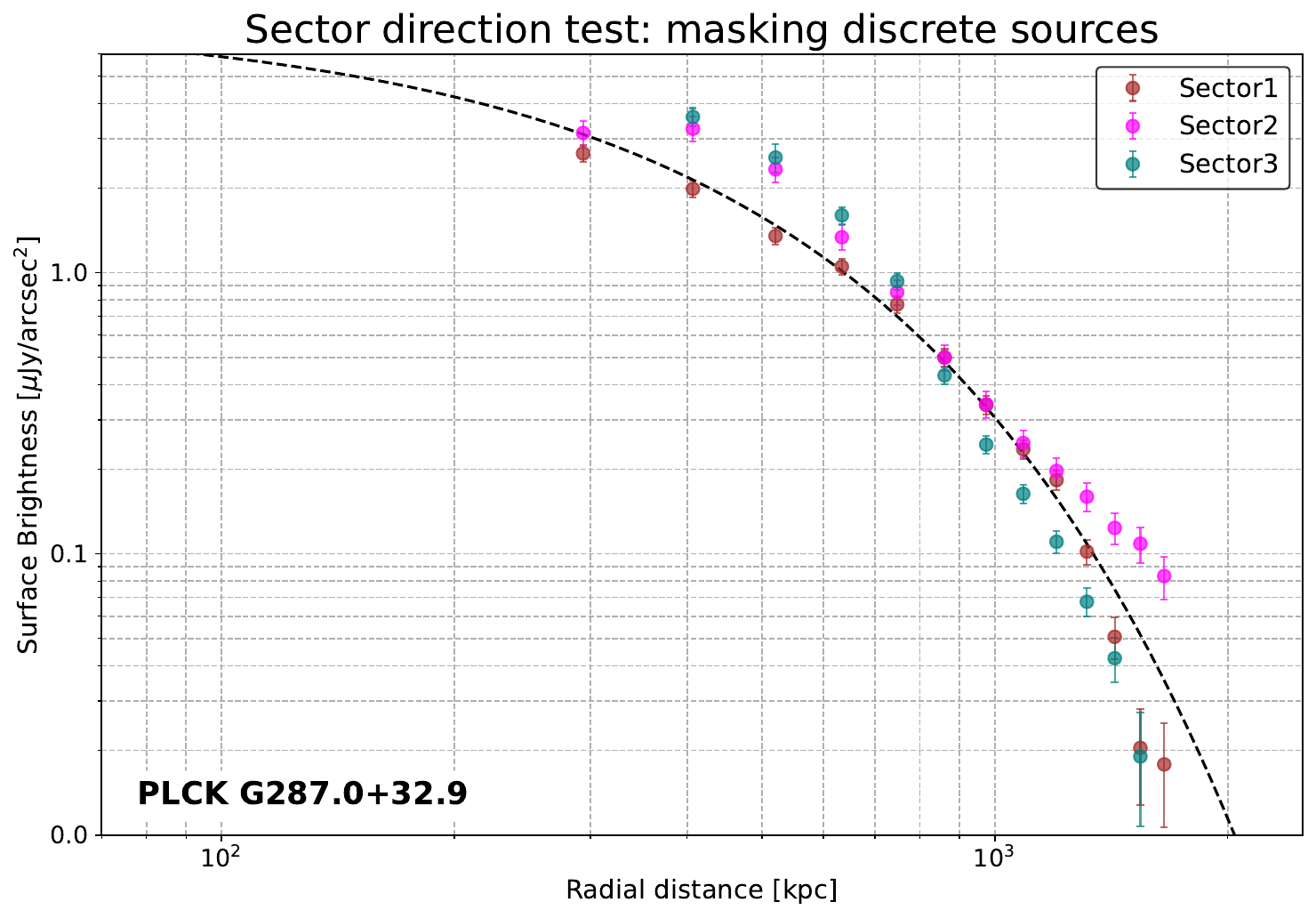}
\includegraphics[width=0.49\textwidth]{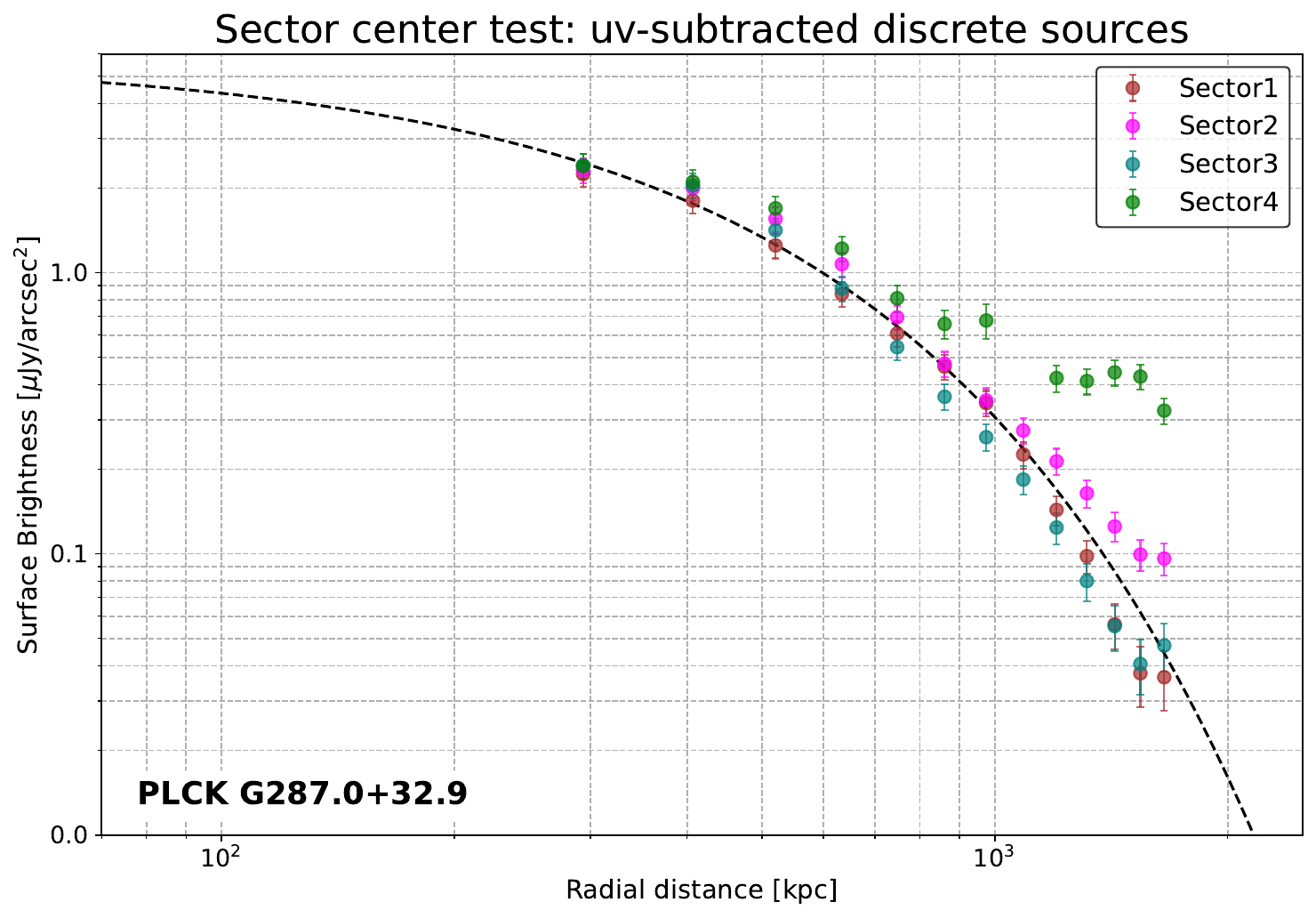}
\includegraphics[width=0.49\textwidth]{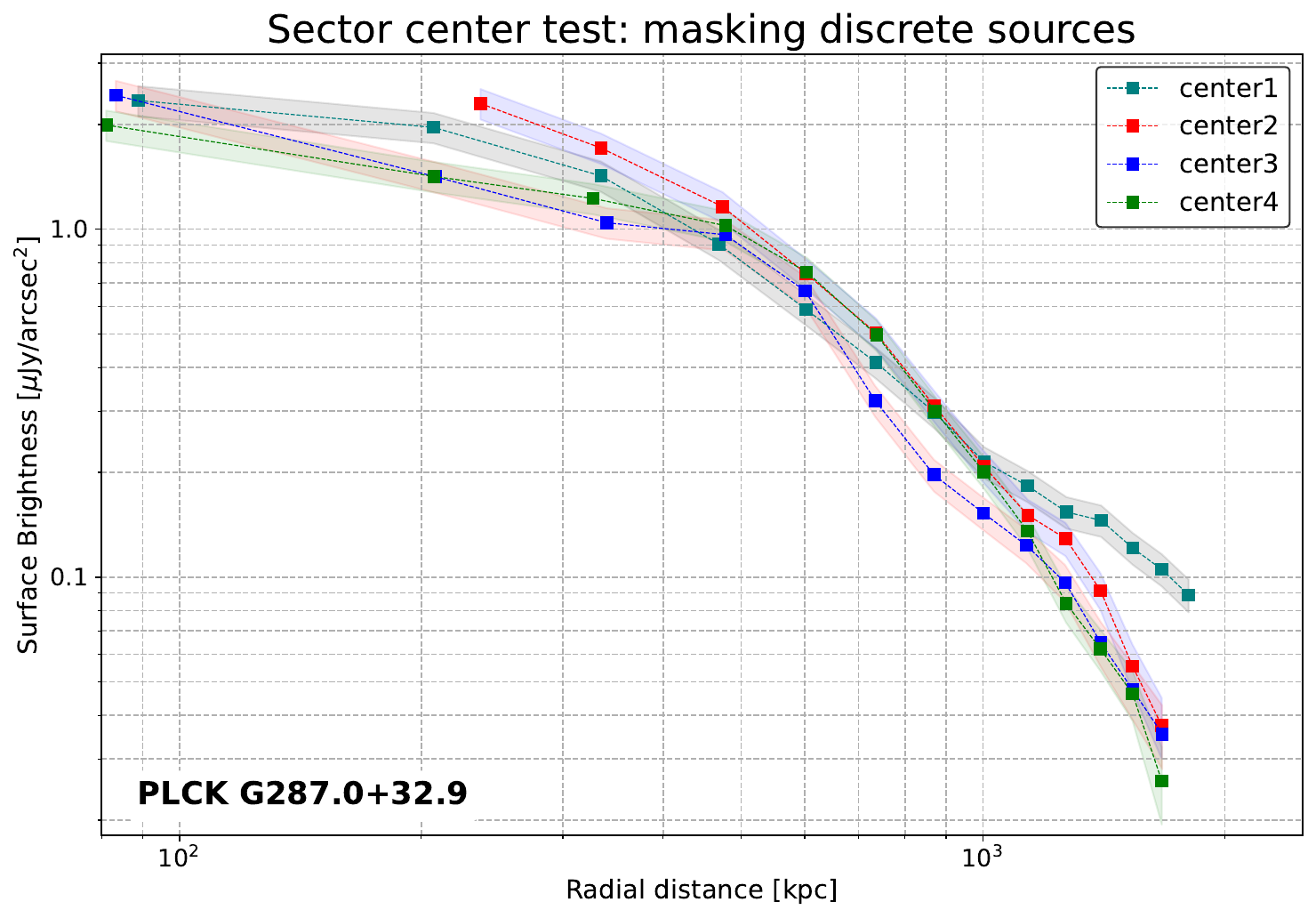}   
\includegraphics[width=0.49\textwidth]{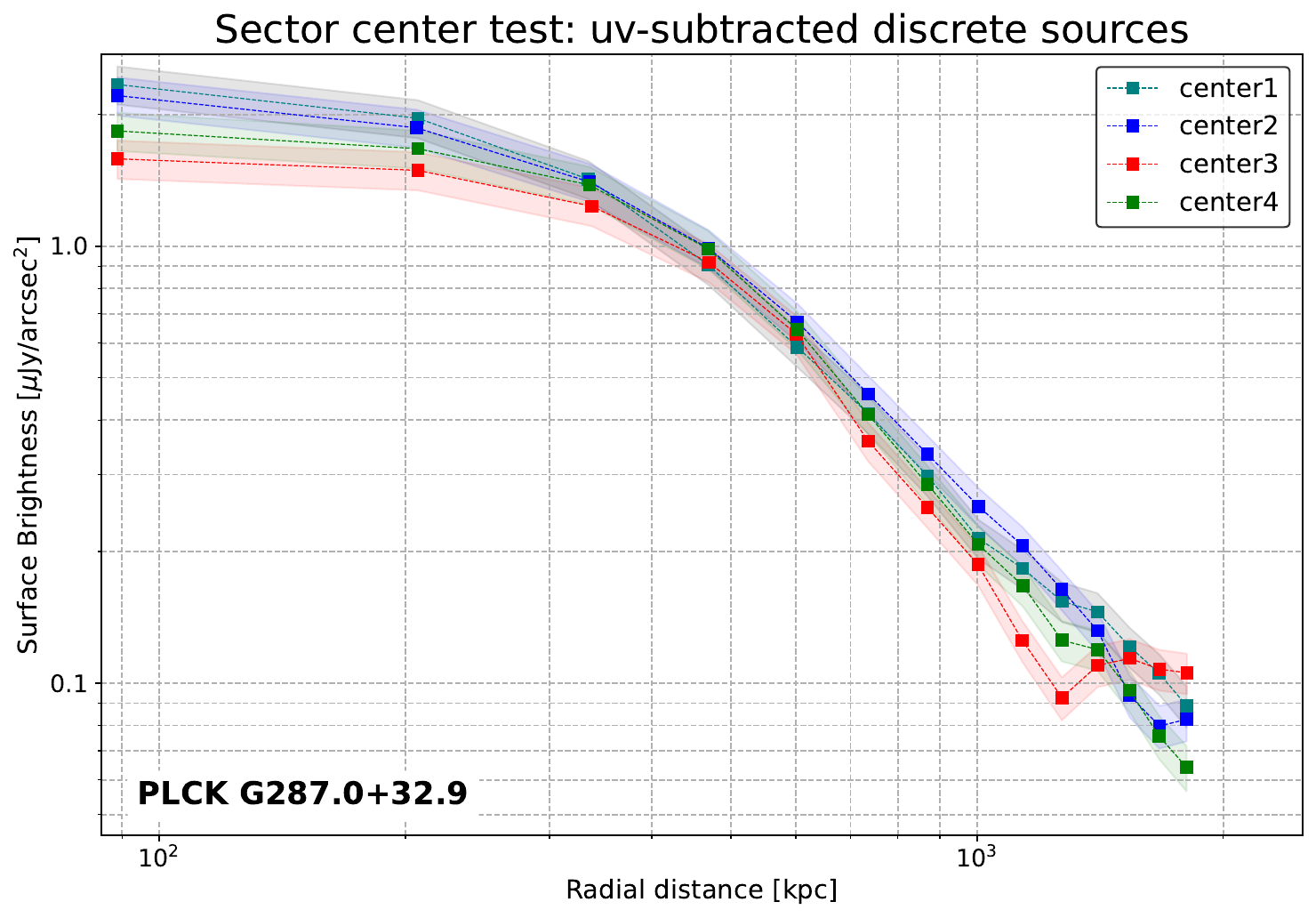}   
\caption{\textit{Top}: Radial profiles extracted using circular,  elliptical, and elliptical partial annuli. Profiles fitted with a single-component exponential model are shown with dashed lines. For the used annuli see Figure\,\ref{fig:grid}. \textit{Middle}: Radial profiles using different sectors; the dashed line shows the fit to the combined profile from all sectors. Sectors are shown in Figure\,\ref{fig::regions} right panel.  \textit{Bottom}: Radial profiles extracted using different choices of centers. The left panels show profiles extracted from an image where unrelated sources within the halo were masked out, while the right panels show profiles obtained from an image where unrelated sources were subtracted from the $uv$-data, as commonly used in the literature. The plot demonstrates that the selection of sectors, sector centers, and discrete source subtraction significantly impacts the extracted radial profiles in radio halos with large extents, potentially introducing an artificial, shallower, second component at larger radii.}
\label{different sectors1}
\end{figure*}

Very recently, \cite{Salunkhe2025} reported uGMRT Band\,3 and Band\,4 observations of PLCK\,G287.0+32.9, claiming the presence of a mega-halo based on two components observed in the radial profile. However, our deeper observations reveal that the second component is not related to the halo. In Figure~\ref{fig::subfeatures}, we present 10\arcsec\ resolution zoom-in images of the northern part of the halo using MeerKAT UHF and uGMRT Band~3 observations. We emphasize that our uGMRT Band~3 observation is a factor of two more sensitive than those reported by \cite{Salunkhe2025} at the same frequency range. Clearly, filamentary structures are embedded within the halo region (marked with curved lines), as seen in our high-frequency, sensitive MeerKAT images. These features are not detected in the Band\,3 images reported by \cite{Salunkhe2025} and are instead included as part of the halo emission. These structures appear more diffuse at lower frequencies and are not associated with the halo emission but rather projected onto it (Rajpurohit et al. in prep). Moreover, the density of discrete sources is higher in our MeerKAT maps compared to the uGMRT Band3 map. As shown in the right panel of Figure~\ref{different sectors}, we also see a second, shallower component in both the 350\,MHz and 815\,MHz profiles when using an image where discrete sources are subtracted from the uv-data (based on the uGMRT Band3 image) as followed by \cite{Salunkhe2025}. However, this second component is introduced by the incomplete subtraction of unrelated sources, in particular, extended structures that are challenging to subtract from the uv-data. When discrete sources are properly masked out, based on sensitive high-frequency images, the radial profile of the PLCK\,G287.0+32.9 halo is well described by a single exponential fit (see Figure~\ref{fig:: PSZ_profiles}) at all the observed frequencies.   

Our analysis suggests that the subtraction/masking of discrete sources from the halo emission is a critical factor when characterizing radial profiles. We emphasize that we carried out this analysis using highly sensitive observations at high frequency. We conclude that the masking process, which uses high-frequency data for compact sources and a combination of high and low frequency data for diffuse unrelated sources (e.g., filaments, relics, tailed radio galaxies) to exclude both thin structures recovered at high frequencies and emission with a steep spectrum detected at low frequencies, is the preferred approach for obtaining reliable radial profiles. If the discrete sources are primarily compact, subtracting them from the $uv$ data provides a reliable intrinsic halo radial profile. However, if extended sources are embedded within the halo emission, the masking approach should be preferred.

\subsection{Annuli and Sector choice}

In the literature, radial profiles are typically extracted using circular or semi-circular annuli. Here, we extend this approach by dividing the halo into subsectors and also employing elliptical annuli to investigate how the choice of sector geometry affects the extracted radial profiles. These subsectors can exhibit different profiles and/or components, reflecting the fact that halos are not entirely circular and often show substructures, as revealed in high-resolution maps of Abell 2255, Abell 2256, Coma, Bullet, MACS\,J0717+3745, and PLCK\,G287.0+32.9  halos \citep{Botteon2022, Rajpurohit2021b, vanWeeren2017b, Sikhosana2023, Bonafede2022, Rajpurohit2023}. 

Our analysis shows significant differences in the resulting profiles when using annuli with different geometries. In the top-left panel of Figure\,\ref{different sectors1}, we present the radial profile of the PLCK\,G287.0+32.9 halo, extracted from an image (with unrelated sources being masked out) using circular, elliptical, and elliptical partial annuli  (see Figure\,\ref{fig:grid}  and \ref{fig::regions} left panel the annuli used) . In the top right panel, the same annuli were used to extract the surface brightness but from an image where unrelated sources were subtracted from the $uv$-data. When directly masking the unrelated sources from the image, we observe only a marginal difference between the shape of the profiles obtained using different annuli and there is no evidence of the second shallower component at larger radii (see Figure\,\ref{different sectors1} top-left panel). However, a clear difference emerges when comparing these profiles to those extracted from a map where unrelated sources were subtracted from the uv data, which shows the presence of a shallower second component (Figure\,\ref{different sectors1} top-right panel).  It is worth noting that in Figure\,\ref{different sectors1} top right panel profile, the northernmost part of the halo is also included. In the middle panels of Figure\,\ref{different sectors1}, we present additional profiles extracted from different sectors. These sectors are shown in Figure\,\ref{fig::regions} left panel. Clearly, the choice of sector impacts the radial profiles regardless of whether a masked or uv-subtracted image is used. This is not surprising, as small sectors can capture substructures that may blend into smoothly varying emission or regions with different properties when using different sectors. In conclusion, just focusing on a particular direction where the halo emission is enhanced may introduce a bias in the radial profiles.

\subsection{Center of the annuli}

To investigate how the selection of the center impacts the extracted profiles, we performed the analysis using multiple center positions and compared the resulting profiles, as shown in Figure\,\ref{different sectors1} bottom panels. The used center positions are shown in Figure\,\ref{fig::regions}. right panel. As evident, the choice of the annuli center introduces a noticeable difference in the profiles, particularly in the outer regions, where deviations become more pronounced.  

When the sector center is chosen at the peak of the radio emission, which is typically close to the X-ray peak, the resulting radial profile typically follows a single exponential fit. However, depending on the sector center, when multiple radio peaks are present in the core region (e.g., Abell 2142 and Bullet) or when discrete sources are embedded in the core region (e.g., PLCK\,G287.0+32.9), the radial profile may show an apparent break at larger radii or deviate from a simple single-component structure. This effect becomes particularly significant when bright or extended discrete sources are located around the cluster core, as their subtraction can strongly affect the choice of the sector center, introducing biases into the radial profile analysis. For example, in PLCK\,G287.0+32.9, three bright compact sources are present in the core region, see Figure\,\ref{fig::PSZ_halo}. Subtracting them from the $uv$-data could shift the location of the radio peak, and thus the sector center, impacting the radial profiles (Figure\,\ref{different sectors1}, bottom panels). 

\section{Conclusion}
We presented new observations of  three massive galaxy clusters,   PLCK\,G287.0+32.9, Abell 2744, and Bullet Cluster conducted with MeerKAT and/or  uGMRT. Our results highlight the emergence of radio halos with large extents (LLS > 2~Mpc) also at high frequencies. In particular, our new images of PLCK\,G287.0+32.9 provide the first high-frequency detection of a radio halo extending out to 3.5~Mpc. 

We analyzed the radial properties of five galaxy clusters, namely PLCK\,G287.0+32.9, Abell 2744, the Bullet Cluster, MACS J0717+3745, and Abell 2142 --- all hosting radio halos. We find that despite their exceptionally large extents, the radio profiles of halos with LLS > 2 Mpc can be described by a single exponential fit, similar to those of classical radio halos. These five clusters are strong candidates for hosting mega-halos --- of which four have been reported to date --- based on their position in the cluster mass versus redshift plot and the total extent of the halo emission.  We find no evidence of a second outer component in their radial profiles. While we were able to obtain profiles with a shallower second component in peripheral regions, it is not real but rather a result of incomplete subtraction of unrelated sources. We emphasize that we used high quality radio data to extract radial profiles. Our findings suggest that radio halos can extend to the cluster periphery, without the transition to an observationally distinguishable different halo component in the outermost regions.

Despite their varying sizes, we find that the radio emissivity of halos with large extents is strikingly similar to that of classical halos, namely $10^{-42}\, \mathrm{erg\,s^{-1}\,cm^{-3}\,Hz^{-1}}$ (except Abell 2142). These halos with large extents lie above the known radio power at 1.4~GHz versus cluster mass scaling relations of halos (except Abell 2142). Moreover, their radial spectral index profiles reveal a clear spectral index gradient from the cluster core to the outer regions. Moreover, the powerful radio halos are found to be more extended, implying that such halos are expected to be detectable out to larger radii ($>$2 Mpc)

Our findings demonstrate that the choice of sectors, annulus centers, and discrete source subtraction significantly affects the radial profiles of halos. Careful subtraction of unrelated sources embedded in the halo is essential to ensure robust and reliable radial profiles. In particular, the identification of multiple components in radio halos based on radial profiles requires caution, as it can result in an artificial second, shallower component due to incomplete subtraction of unrelated sources and the choice of annuli/sectors. Additional observational properties such as radial spectral index profiles and radio versus X-ray comparisons should be considered to assess the presence of a secondary component \citep[e.g.,][]{Botteon2020, Rajpurohit2021c, Rajpurohit2021b, Rajpurohit2022b, Bruno2023, Biava2024}.

In conclusion, we find that halos with large extents ($>2$\,Mpc) share key characteristics with classical radio halos: 1) their averaged radial profiles are well described by a single-component exponential fit, 2) they show radial spectral index steepening, and 3) radio emissivity of $\sim 10^{-42}\, \mathrm{erg\,s^{-1}\,cm^{-3}\,Hz^{-1}}$. Our results highlight that the distinction between different categories of radio halos, such as hybrid halos, giant halos, and mega-halos, based on size is becoming increasingly blurred. In contrast to the four mega halos with $\rm LLS> 2\,Mpc$, the five systems analyzed in this study do not exhibit different properties than those generally found in radio halos. Our results demonstrate that the observable size of a halo is primarily determined by the image depth and uv-coverage relative to its central brightness and e-folding radius. The $>2$~Mpc extents observed in some cases reflect the presence of bright, powerful radio halos, allowing their emission to be detected out to larger radii.

\section*{Acknowledgments}
KR acknowledges funding support from the Smithsonian Combined Support for Life on a Sustainable Planet, Science, and Research administered by the Office of the Under Secretary for Science and Research and NASA grant 619246. WF acknowledges support from the Smithsonian Institution, the Chandra High Resolution Camera Project through NASA contract NAS8-03060, and NASA Grants 80NSSC19K0116 and GO1-22132X. CJ acknowledges support from the Smithsonian Institution. LB acknowledges support from the NextGenerationEU funds within the National Recovery and Resilience Plan (PNRR), Mission 4 - Education and Research, Component 2 -From Research to Business (M4C2), Investment Line 3.1- Strengthening and creation of Research Infrastructures, Project IR0000026- Next Generation Croce del Nord. The research leading to these results has received funding from the European Union's Horizon 2020 research and innovation programme under grant agreement No 101004719 [ORP]. MeerKAT data reduction were conducted on the Smithsonian High Performance Cluster (SI/HPC), Smithsonian Institution https://doi.org/10.25572/SIHPC. The authors thank the staff of the MeerKAT observatory for their help with the observations presented in this work.  The MeerKAT telescope is operated by the South African Radio Astronomy Observatory, which is a facility of the National Research Foundation, an agency of the Department of Science and Innovation. We thank the staff of the GMRT that made these observations possible. GMRT is run by the National Centre for Radio Astrophysics of the Tata Institute of Fundamental Research. This research made use of the LOFAR-IT computing infrastructure supported and operated by INAF, including the resources within the PLEIADI special ``LOFAR'' project by USC-C of INAF, and by the Physics Dept. of Turin University (under the agreement with Consorzio Interuniversitario per la Fisica Spaziale) at the C3S Supercomputing Centre, Italy.

\facilities{MeerKAT}, {GMRT}, {LOFAR}

\software{CARACal \citep{caracal2020}, AOflagger \citep{Offringa2010}, WSClean \citep{Offringa2014}, SPAM \citep{Intema2009}, APLpy \citep{aplpy}, Matplotlib \citep{matplotlib}}

\bibliographystyle{aasjournal}
\bibliography{ref}

\end{document}